\documentclass[12pt,letter]{article}
\usepackage{graphicx,subfigure}
\usepackage{epstopdf}
\usepackage{epsfig,amssymb}
mm\evensidemargin=-5true mm \topmargin=-15true mm

\newcommand{\beq}{\begin{equation}}
\newcommand{\eeq}{\end{equation}}
\newcommand{\be}{\begin{equation}}
\newcommand{\ee}{\end{equation}}
\newcommand{\beqa}{\begin{eqnarray}}
\newcommand{\eeqa}{\end{eqnarray}}
\newcommand{\beqar}{\begin{eqnarray*}}
\newcommand{\eeqar}{\end{eqnarray*}}
\newcommand{\bea}{\begin{eqnarray}}
\newcommand{\eea}{\end{eqnarray}}

\newcommand{\s}{\sigma}

\newcommand{\n}{\nabla }

\begin{document}

\setlength{\unitlength}{1mm}

\begin{titlepage}

\rightline{}

\begin{flushright}
\end{flushright}
\vspace{1cm}

\begin{center}
{\bf \Large Covariant Counterterms and Conserved Charges in Asymptotically Flat Spacetimes}
\end{center}

\vspace{1cm}

\begin{center}
Robert B. Mann$^{a}$, Donald Marolf$^{b}$,  and Amitabh
Virmani$^{b}$

\vspace{1cm}{\small {\textit{$^{a}$Dept. of Physics, University of
Waterloo, Waterloo, Ontario N2L 3G1, Canada \\
and Perimeter Institute
for Theoretical Physics, Ontario N2J 2W9, Canada }}}\\
\vspace{2mm} {\small {\textit{$^{b}$Department of Physics,
University of California, Santa Barbara, CA 93106-9530,
USA}}}\\
\vspace*{0.5cm} {\tt rbmann@sciborg.uwaterloo.ca,
marolf@physics.ucsb.edu,
 virmani@physics.ucsb.edu}
\end{center}

\vspace{0.5cm}

\begin{center}
\textit{ Dedicated to Rafael Sorkin on the occasion of his 60th birthday.}
\end{center}

\vspace{0.5cm}

\begin{abstract}
Recent work has shown that the addition of an appropriate covariant boundary term to the gravitational action yields a well-defined variational principle for asymptotically flat spacetimes and thus leads to a natural definition of conserved quantities at spatial infinity.  Here we connect such results to other formalisms by showing explicitly i) that for spacetime dimension $d \ge 4$ the canonical form of the above-mentioned covariant action is precisely the ADM action, with the familiar ADM boundary terms and ii) that for $d=4$ the conserved quantities defined by counter-term methods agree precisely with the Ashtekar-Hansen conserved charges at spatial infinity.
\end{abstract}

\end{titlepage}


\vspace{0.5cm} \tableofcontents


\setcounter{equation}{0}


\section{Introduction}
\label{introduction}

The fact that there is no local stress tensor for the
gravitational field has encouraged a variety of definitions of
conserved quantities corresponding to the Poincar\'e symmetries of
asymptotically flat spacetimes \cite{ADM1}-\cite{MM}. A useful
perspective on this issue was gained by Arnowitt, Deser, and
Misner (ADM) \cite{ADM1} by  focussing on the gravitational
Hamiltonian.  As emphasized by Regge and Teitelboim \cite{RT}, one
attains an essentially unique result by demanding that the
conserved charges be well-defined Hamiltonian generators of the
associated asymptotic symmetries.  In particular, due to
diffeomorphism-invariance the non-trivial part of such generators
is just a boundary term, and the boundary terms are determined by
the requirement that the charges generate the symmetries.

It has long been known that, with a well-defined Hamiltonian $H$
in hand, it is straightforward to construct a canonical action $S
= \int \pi \dot{g} - H$ for general relativity.  Furthermore, the
boundary terms in $H$ guarantee that this action is stationary on
appropriately asymptotically flat solutions and thus provides a
valid variational principle for this context.  However,  covariant
and background-independent action principles with this property
have only recently been demonstrated \cite{eeA,ABL,MM}; the
proposals of \cite{Mann, KLS} were also shown to provide valid
variational principles in \cite{MM}. We will focus on the
treatment of \cite{MM}, which uses a second-order formulation in
terms of metric variables. The key to the construction of this
action principle is the addition of an appropriate boundary term
(called a ``counter-term") to the Einstein-Hilbert action with
Gibbons-Hawking term.

Given a well-defined diffeomorphism-invariant variational
principle, conserved generators of asymptotic symmetries are
readily constructed (see e.g.\cite{HIM2}) using the covariant
phase space as embodied in the Peierls bracket \cite{Peierls}.
Such a construction was performed in \cite{MM} for asymptotically
flat spacetimes, where it was argued on general grounds that
conserved quantities defined in this way must agree with the older
definitions \cite{ADM1}-\cite{MM}.
 It was also explicitly shown that such Peierls
definitions of energy and momentum at spatial infinity agree with
that of Ashtekar and Hansen \cite{AshtekarHansen}.  The key point
here is that the Peierls conserved quantities can be expressed
\cite{HIM2} through a boundary stress tensor $T_{ab}$, similar to
that used  in the quasi-local context \cite{BY} or in anti-de
Sitter space (see e.g. \cite{skenderis,Kraus}).  In particular,
the conserved quantity associated with an asymptotic Killing field
$\xi^a$ is given by the flux of the current $T_{ab} \xi^a$ through
a cut of spatial infinity.   As we will review below, the
Ashtekar-Hansen definition of momentum and energy is very similar,
but in terms of the current $E_{ab} \xi^a$ built from the electric
part $E_{ab}$ of the Weyl tensor.  Equality of these definitions
then follows from the fact that the leading term in $T_{ab}$ turns
out to be just $E_{ab}$ times a normalization factor.  A similar
relationship holds in the asymptotically anti-de Sitter context
\cite{HIM,KSnew}.

The purpose of the present work is to exhibit in more detail the
relationship between the counter-term techniques of \cite{MM} and
the previous literature.  We exhibit two main results.  We first
perform a canonical (space + time) reduction of the covariant
action of \cite{MM}  for spacetime dimension $d \ge 4$ and show
that it yields {\it precisely} the ADM action, with the familiar
ADM boundary terms.  This allows one to see explicitly that all
boundary stress tensor charges of \cite{MM} agree with the charges
defined by ADM \cite{ADM1}.  Corresponding results in the $d=4$
first order formalism were derived in \cite{eeA,ABL} for the
boundary term introduced there. Note that, in contrast, for the
case $d=3$ the natural covariant action differs from the canonical
action of \cite{AV} by a term which shifts the zero of the
Hamiltonian, corresponding to the assignment of a non-zero energy
to 2+1 Minkowski space \cite{DonLeo}.

The second result follows up on the observation that the boundary stress tensor momentum and energy are explicitly equal to the Ashtekar-Hansen definitions of these charges.    Here we investigate the relationship between the corresponding definitions of Lorentz generators.   This calculation is interesting because the boundary stress tensor definition is naturally thought of in terms of the sub-leading behavior of the electric part of the Weyl tensor, while the Ashtekar-Hansen definition uses the magnetic part of the Weyl tensor.  Nonetheless, we are able to explicitly demonstrate agreement by making sufficient use of the equations of motion.

The plan of this paper is as follows.  We begin with various definitions and a brief review of the covariant counterterm construction of \cite{MM} in section
\ref{review}.  We then perform the canonical reduction of the resulting covariant action in section  \ref{section.ADM}, obtaining the ADM action as a result.  In section \ref{covariant},  we show explicit agreement between the boundary stress tensor charges of \cite{MM} and the Ashtekar-Hansen charges \cite{AshtekarHansen}, focussing on the case of Lorentz generators (boosts and rotations).    Here, a
number of calculational details are relegated to appendices
\ref{pi}, \ref{sectionidentity} and \ref{collection}.
Of course, the agreement between the two sets of covariant charges follows from the agreement between each set of expressions with those of ADM.  However, it is instructive to show the agreement directly, without passing through the canonical formalism; i.e., without decomposing spacetime into separate notions of space and time.  This agreement also provides a check of our earlier results.  We close with some discussion in section \ref{disc}.

\setcounter{equation}{0}


\section{Preliminaries: Asymptotic Flatness, Actions, and Conserved Charges}
\label{review}

Here we set the stage for our later work by providing relevant definitions and review.  After stating our notion of asymptotic flatness, we introduce the variational principle of \cite{MM} and the associated definition of conserved charges.

\subsection{Asymptotic Flatness}

We begin with the definition of asymptotic flatness from \cite{MM}, which was directly inspired by that of \cite{ABR}.    We will give a coordinate-based definition here, modeled on the definitions of \cite{BS,B}.  The results are readily translated to the geometric language of either the Spi formalism \cite{AshtekarHansen,AshtekarInf} or  that of \cite{AR}.  The definition of asymptotic flatness given below is particularly close to that of \cite{AR}, which treats spatial infinity as the unit time-like hyperboloid.

We are interested in $d \ge 4$ spacetimes which are asymptotically flat at spatial infinity in the sense that the line
element admits an expansion of the form
\begin{equation}  \label{AFdef}
ds^2 = \left( 1+ \frac{2 \sigma}{\rho^{d-3}} + \mathcal{O}(\rho^{-(d-2)})%
\right) d\rho^2 + \rho^2 \left( h^0_{ij} + \frac{h^1_{ij}}{\rho^{d-3}}
+ \mathcal{O}(\rho^{-(d-2)}) \right) d\eta^i d\eta^j +  \rho \left( \mathcal{O}(\rho^{-(d-2)}) \right) d\rho d\eta^j,
\end{equation}
for large positive $\rho$. Here, $h^0_{ij}$ and $\eta^i$ are a
metric and the associated coordinates on the unit $(d-2,1)$
hyperboloid ${\cal H}^{d-1}$ (i.e., on $d-1$ dimensional de Sitter
space) and $\sigma, h^1_{ij}$ are respectively a smooth function
and a smooth tensor field on ${\cal H}^{d-1}$. Thus, $\rho$ is the
``radial'' function
associated with some asymptotically Minkowski coordinates $x^a$ through $%
\rho^2 = \eta_{ab} x^a x^b.$ In (\ref{AFdef}), the symbols $\mathcal{O}%
(\rho^{-(d-2)})$ refer to terms that fall-off at least as fast as $%
\rho^{-(d-2)}$ as one approaches \textit{spacelike} infinity, i.e., $\rho
\rightarrow +\infty$ with fixed $\eta$.
The inclusion of NUT-charge requires some changes in the global structure, but these changes have little effect on the arguments below.
Note that, for $d=4$, any metric that is asymptotically flat at spatial infinity by
the criteria of any of
\cite{AshtekarHansen,AR,Wald, ABR} also satisfies
(\ref{AFdef}).    In $d\ge 5$ dimensions, the definition (\ref{AFdef}) is more
restrictive than that of \cite{skenAF}, which for $d \ge 5$
allows additional terms of order $\rho^{-k}$ for $d-4 \ge k \ge 1$
relative to the leading terms.  However, (\ref{AFdef}) is at
least as general as the definition which would result by applying
the methods of \cite{RT}; i.e., by considering the action of the
Poincar\'e group on the Schwarzschild spacetime.

Below, we consider vacuum solutions or, what is equivalent for our
purposes, solutions for which the matter fields fall off
sufficiently fast at infinity that the leading order contribution
to $R_{ikjl}$ comes only from the Weyl tensor (i.e., the
contribution at that order from the Ricci tensor vanishes).  This
is the case for typical configurations of matter fields.
In an asymptotically Cartesian frame, finiteness of the matter
stress-energy (i.e., of $\int T_{00} \sqrt{g_\Sigma}$) requires
$T_{00} \sim \rho^{-(d-1 + \epsilon)}$ for some $\epsilon > 0$.  But
then $R_{ij}$ falls off at a similar rate, while we consider the
term in $R_{ijkl}$ of order just $\rho^{-(d-1)}$.  That is, if the
matter fields fell off sufficiently slowly so as to contribute at the same order as the Weyl tensor, then the total energy would diverge
logarithmically.

\subsection{A Variational Principle}
\label{varprinc}

Variations of the Einstein-Hilbert action with Gibbons-Hawking term preserving (\ref{AFdef}) reduce to a boundary term, but this term does not vanish for generic asymptotically flat variations.  In \cite{MM} it was found that
a fully stationary action could be obtained by adding an additional boundary term.  We therefore follow
\cite{MM} in considering  the action
\begin{equation}
\label{covaction} S = \frac{1}{16\pi G} \int_{\cal M} \sqrt{-g}R +
\frac{1}{8\pi G} \int_{\partial {\cal M}} \sqrt{-h} (K - \hat K),
\end{equation}
where $\hat K := h^{ij} \hat K_{ij}$ and $\hat K_{ij}$ is defined
to satisfy
\begin{equation}
\label{Khat} {\cal R}_{ij} = \hat K_{ij} \hat K - \hat K_i^m \hat
K_{mj},
\end{equation}
where ${\cal R}_{ij}$ is the Ricci tensor of the boundary metric
$h_{ij}$ on $\partial {\cal M}$ and we follow the conventions of
Wald \cite{Wald}.
In solving (\ref{Khat}), we choose the solution of (\ref{Khat}) that asymptotes to the extrinsic curvature of the boundary of Minkowski space as $\partial {\cal M}$ is taken to infinity.
As described in \cite{MM}, the boundary term (\ref{Khat}) can be motivated, via
the Gauss-Codazzi equations, from the heuristic idea (see e.g. \cite{HH, BY,BCM}) that one should
subtract off a ``background" divergence.

The boundary terms above are defined by a limiting procedure in which one considers a  one-parameter
family of regions $ \mathcal{M}_\Omega  \subset \mathcal{M}$ which form an increasing family converging to $%
\mathcal{M}$. Any such family represents a particular way of `cutting off' the
spacetime $\mathcal{M}$ and then removing this cut-off as $\Omega
\rightarrow \infty$. Thus, all terms in the action (\ref{covaction})
are to be understood as the $\Omega \rightarrow \infty$
limits of families of functionals  in which $(\mathcal{M}, {%
\partial \mathcal{M}}, {\ h})$ are replaced by $(\mathcal{M}_\Omega, {%
\partial \mathcal{M}}_\Omega, {h}_\Omega)$. We will take this cut-off to be
specified by some given function $\Omega$ on $\mathcal{M}$ such that $\Omega
\rightarrow \infty$ at spatial infinity. We define $\mathcal{M}_{\Omega_0}$
to be the region of $\mathcal{M}$ in which $\Omega < \Omega_0$, so that $%
(\partial \mathcal{M}_{\Omega_0}, h_{\Omega_0})$
is the hypersurface where $\Omega = \Omega_0$.

We will be interested in two distinct notions of the boundary $\partial {\cal M}$.
The first notion preserves manifest Lorentz invariance.   To this end,  one may consider
the class of ``hyperbolic cut-offs,'' in which
$\Omega$ is taken to be some function of the form:
\begin{equation}  \label{hypcut}
\Omega^{hyp} = \rho + O(\rho^{0}).
\end{equation}
Choosing such a cut-off leads to a hyperbolic representation of spacelike
infinity directly analogous to the construction of Ashtekar and Romano \cite{AR}.
The cut-off in time is also important, and we take this to be given by initial and final Cauchy surfaces $\Sigma_-, \Sigma_+$.  In the work below we will always associate $\Omega^{hyp}$ with $\Sigma_\pm$ which asymptote to fixed Cauchy surfaces $C_+$ and $C_-$ of the hyperboloid ${\cal
H}^{d-1}$; that is, we allow $\Sigma_\pm$ to be defined by any
equations of the form
\begin{equation}
0 =  f_\pm (\eta) + {\cal O}(\rho^{-1}),
\end{equation}
for smooth functions $f_\pm$ on ${\cal H}^{d-1}$.  One may think
of such surfaces $\Sigma_-,\Sigma_+$ as being locally boosted
relative to each other at infinity.   Note that when ${\cal M}$ is defined by such
past and future boundaries the volume of $\partial
\mathcal{M}_\Omega$ grows as $\rho^{d-1}$.

While the hyperbolic cut-off is a natural choice for covariant investigations, the connection with the canonical formalism is more natural when one uses a ``cylindrical'' representation of $\partial {\cal M}$, associated with the family of cut-off functions
\begin{equation}  \label{cylcut}
\Omega^{cyl} = r + O(\rho^{0}).
\end{equation}
In (\ref{cylcut}), the coordinate $r$ is defined by $r^2 = \rho^2 + t^2$ and
$t$ is an asymptotically Minkowski time coordinate. More precisely, we may
define $t$ through the requirement that the metric (\ref{AFdef}) takes the
form
\begin{eqnarray}  \label{cyldef}
ds^2 = - \left( 1+ \mathcal{O}(\rho^{-(d-3)})\right) dt^2 + \left( 1+
\mathcal{O}(\rho^{-(d-3)})\right) dr^2 + \cr r^2 \left( \omega_{IJ} +\mathcal{O}%
(\rho^{-(d-3)}) \right) d\theta^I d\theta^J  + \mathcal{O}%
(r^{-(d-4)})  d\theta^I dt  ,
\end{eqnarray}
where $\omega_{IJ}, \theta^I$ are the metric and coordinates on the unit $%
(d-2)$-sphere.

We may now state the main conclusions of \cite{MM}.
The action (\ref{covaction}) defined using the cylindrical cut-offs\footnote{As we note in section \ref{covariant} below, the definition of the counter-term requires a slight refinement when defined by a cylindrical
cut-off for $d=4$.}
 above is stationary under any asymptotically flat variation (i.e., preserving (\ref{AFdef})) about a solution to the equations of motion for $d \ge 4$.  In addition, when defined using the hyperbolic cut-offs above, the action
 is stationary under any asymptotically flat variation (i.e., preserving (\ref{AFdef})) about a solution to the equations of motion for $d \ge 5$.   The case $d=4$ with hyperbolic cut-off is somewhat special.  In that case, the action is stationary only under those variations which satisfy
\begin{equation}
\label{varh1} \delta h^1_{ij} = \alpha h^0_{ij},
\end{equation}
 for $\alpha$ a smooth function on ${\cal H}^{3}$.
 This restriction may be justified as the restriction of the domain of the action (\ref{covaction}) to a single covariant phase space of the sort defined in \cite{ABR}.
 As implied by (\ref{hypcut}) and (\ref{cylcut}), the action (\ref{covaction}) will depend only on the asymptotic form of $\Omega$, which we
will take to represent a fixed auxiliary structure.
As usual, the case  $d=3$ must be handled separately \cite{DonLeo}.


\subsection{Boundary Stress Tensors and Conserved Charges}
\label{bst}

We now review the boundary stress tensor construction of conserved
charges from a variational principle.   The idea of a boundary
stress tensor for asymptotically flat spacetimes was introduced in
\cite{BY}, and a covariant version was described in \cite{Mann2}
following \cite{skenderis,Kraus}.   This version was used in
\cite{DAR} to compute conserved quantities for black rings
and   in \cite{KKmonopole} to calculate  the mass of the Kaluza-Klein magnetic monopole. In \cite{MM}, the arguments of \cite{HIM2}
were adapted to show that the resulting conserved charges generate
the expected asymptotic symmetries via the Peierls bracket
\cite{Peierls}.

At the operational level, the boundary stress tensor is straightforward to introduce.  Consider the family of actions $S_\Omega$ associated with the family ${\cal M}_\Omega$ of regulated spacetimes for any cutoff function $\Omega$.  For each $\Omega$, define the boundary stress tensor
\begin{eqnarray}  \label{Texp}
T_{ij} (\Omega)&:=& \frac{-2}{\sqrt{-h}} \frac{\delta S_{\Omega} }{%
\delta h_\Omega^{ij}}  \cr %
&:=& \Omega^{-(d-4)} \left( T_{ij}^{0} + \Omega^{-1} T_{ij}^1 + \mbox{ terms
vanishing  faster  than} \ \Omega^{-1} \right).
\end{eqnarray}
Here the variations are taken with respect to metric  components
$h_{ab}$ on the boundary and are computed about solutions to the
equations of motion, so that there is no contribution from the
interior.

For any asymptotic symmetry $\xi$, one defines
 the charge
\begin{equation}
\label{Qhyp}
Q[\xi] =
 \int_{C}  \sqrt{h_{C}}  T_{ij} \xi^i n^j,
\end{equation}
where the integral is over some Cauchy surface $C$ of the hyperboloid ${\cal H}^{d-1}$.
This definition is sufficient for our present purposes, though additional terms may contribute in more general situations \cite{HIM2}.
It was shown in \cite{MM} that (\ref{Qhyp}) is independent of the particular choice of $C$, except perhaps for the case where $\xi$ contains an asymptotic boost and one has used the cylindrical cutoff $\Omega^{cyl}$.  The results of section \ref{section.ADM} below will establish that (\ref{Qhyp}) is independent of $C$ for this case as well.

Let us briefly compare (\ref{Qhyp}) with two other definitions of conserved quantities.
We first note that (\ref{Qhyp}) strongly resembles the expressions for asymptotically flat energy and momentum introduced by Ashtekar and Hansen \cite{AshtekarHansen}.  In particular, it was pointed out in \cite{MM} that when $\xi$ is an asymptotic translation, only the leading term in $T_{ij}$ contributes to (\ref{Qhyp}).  Furthermore, when the boundary $\partial {\cal M}$ is defined by a hyperbolic cut-off (\ref{hypcut}) it was shown by direct computation in \cite{MM} that
\begin{equation}
\label{TandE}
T_{ij} = \frac{1}{8\pi G} \frac{\rho^{-(d-4)} }{d-3}E_{ij} + {\cal O} (\rho^{-(d-3)}),
\end{equation}
where $E_{ij}$ is the pull-back to the hyperboloid of $E_{ac} =
\rho^{d-3} C_{abcd} \rho^b \rho^d$; i.e., $E_{ij}$ is the first non-trivial
term in the electric part of the Weyl tensor\footnote{In \cite{MM}, the symbol $E_{ac}$ was used to denote the full
electric Weyl tensor, and so did not include the factor of $\rho^{d-3}$.}. Here  $\rho^a$ is the
unit normal to the hyperboloid at constant $\rho$ and we have used
the fact that the vacuum Einstein equations hold to leading
non-trivial order. As a result, for such $\xi$ one may rewrite
(\ref{Qhyp}) as
\begin{equation}
\label{QAH}
Q[\xi] = \frac{1}{8\pi G (d-3)}
 \int_{C}  \sqrt{ h^0_{C}}  E_{ij}   (\xi^0)^i (n^0)^j,
\end{equation}
where $h^0_C$ is the determinant of the metric induced on the cut $C$ by metric on the {\it unit} hyperboloid ($ h^0_{ij} = \rho^{-2} h_{ij}+ \dots $), $(n^0)^j = \rho n^j$ is the unit normal to $C$ with respect to $ h^0_{ij}$, and we have introduced the corresponding $(\xi^0)^i = \rho \xi^i$.  All of the factors in (\ref{QAH}) are now normalized so that they are independent of $\rho$ at large $\rho$ (and (\ref{QAH}) is manifestly finite).  For $d=4$, the expression (\ref{QAH}) is exactly the Ashtekar-Hansen definition of the conserved charges corresponding to translations.  This direct relationship motivates us to ask in section \ref{covariant} if there is a corresponding relationship between the  Lorentz charges (i.e., angular momentum and boost generators).  We will see there that the relationship is significantly more complicated.

Finally, it is interesting to compare (\ref{Qhyp}) with the ADM definitions \cite{ADM1} of these charges.
As noted above, the definition of $Q[\xi]$ depends on the choice of an action.  It is perhaps reassuring to note that using the canonical ADM form of the action to calculate (\ref{Qhyp}) results in charges $Q[\xi]$ which are explicitly the familiar ADM definitions for asymptotically Poincar\'e transformations.  To see this, consider the canonical action
\begin{equation}
\label{can} S = \int \left( \tilde \pi^{ab} \dot{q}_{ab} - N
\tilde {\cal H} - N^a \tilde {\cal H}_a \right) - \frac{1}{16 \pi
G} \int_{\partial M}  \left(N     {\cal E}^{ADM} + N^a   {\cal P}_a^{ADM} \right).
\end{equation}
Here, $q_{ab}= g_{ab} + n_a n_b$, $N$, $N^a$, and $n^a$ are the usual spatial metric, lapse, shift, and unit future-directed timelike normal associated with the foliation of spacetime implicit in (\ref{can}) above (see e.g. \cite{ADM1,Wald}).   The dot represents the Lie derivative with respect to the vector field
$t^a = Nn^a + N^a$, with tensor indices pulled back into the sheets of the foliation.

Although we introduced the boundary stress tensor $T_{ij}$ as a tensor with indices in the co-tangent space of the boundary manifold $\partial {\cal M}$, we can use the natural pull-back to define $T_{ab}$, with indices in the co-tangent space of the bulk spacetime.  Note that we have
\begin{equation}
T_{ab} = - \frac{2}{\sqrt{-h}} \frac{\delta S}{\delta h^{ab}} = -
\frac{2}{\sqrt{-h}} \frac{\delta S}{\delta g^{ab}} ,
\end{equation}
since the variation of (\ref{can}) about a solution involves only $\delta h^{ab}$.

It is clear that the ADM conserved quantities take a simple form
in terms of the variations $\frac{\delta S_\Omega}{\delta
N_\Omega}$ and $\frac{\delta S_\Omega}{\delta N^a_\Omega}$, where
as in (\ref{Texp}) the variations are taken with respect to metric
components on the boundary and are  computed about solutions to
the equations of motion.  We may relate such variations to those
in (\ref{Texp}) by noting that
\begin{equation}
\frac{\delta g^{ab}}{\delta N}  = \frac{\delta}{\delta N} \left( -
n_a n_b \right)  =  \frac{2}{N} n_a n_b,
\end{equation}
and
\begin{equation}
\frac{\delta g^{ab}}{\delta N^c}  = \frac{\delta}{\delta N^c} \left( -
n^a n^b \right)  =   \frac{1}{N} (n^a \delta^b_c + n^b \delta^a_c).
\end{equation}

Let us choose our slice such that the unit normal is an asymptotic time translation.  Then we may compute the energy by taking $\xi =n$,  for which we have
\begin{eqnarray}
\label{Qt}
Q[n] &=& \int_{C} \sqrt{h_C} n^a n^b T_{ab} = - \int_{C}
\sqrt{h_C} \frac{2n^a n^b}{ \sqrt{-h }} \frac{\delta
S}{\delta g^{ab}}\cr &=& - \int_{C} \frac{\delta g^{ab}}{\delta N}
\frac{\delta S}{\delta g^{ab}} = - \int_{C}  \frac{\delta S}{\delta
N} =  \frac{1}{16\pi G}  \int_{C} {\cal E}_{ADM}.
\end{eqnarray}
 We see that applying the boundary stress tensor construction
of energy to (\ref{can}) yields precisely the ADM energy.
Similarly, for $\xi^a n_a =0$ we have
\begin{eqnarray}
\label{Qspace}
Q [\xi] &=& \int_{C} \sqrt{h_C} n^a \xi^b T_{ab} =
- \int_{C} \sqrt{h_C} \frac{2n^a\xi^b}{ \sqrt{-h }}
\frac{\delta S}{\delta g^{ab}}\cr &=& - \int_{C} \xi^c
\frac{\delta g^{ab}}{\delta N^c} \frac{\delta S}{\delta g^{ab}} =
-\int_{C} \xi^a \frac{\delta S}{\delta N^a} = \int_{C} \xi^a {\cal P}^{ADM}_a
 \cr &=& Q_{ADM}[\xi] .
\end{eqnarray}
Finally, the corresponding result for boosts follows by taking a linear combination of (\ref{Qt}), (\ref{Qspace}).


\setcounter{equation}{0}
\section{Equivalence of the $\hat K$ and ADM actions}
\label{section.ADM}

This section performs the space + time reduction of the covariant action (\ref{covaction}) with $\hat K$ counter-term in any spacetime dimension $d \ge 4$, using the ``cylindrical'' definition of the boundary terms as described in section \ref{varprinc}.  As a result, $h_{ij}$ becomes asymptotically the metric on the standard cylinder of radius $r$:
\begin{equation}  \label{hcyl}
h_{ij} dx^i dx^j  = - \left( 1+ \mathcal{O}(r^{-(d-3)})\right) dt^2 + r^2 \left( \omega_{IJ} +\mathcal{O}%
(r^{-(d-3)}) \right) d\theta^I d\theta^J + \mathcal{O}%
(r^{-(d-4)})  d\theta^I dt  .
\end{equation}

We will show that this reduction leads to the familiar ADM action (\ref{can}), including precisely the ADM boundary terms.  Thus, our goal is very similar to that of \cite{HH}, though we consider the $\hat {K}$ boundary term instead of a boundary term defined by background subtraction\footnote{The definition of the $\hat K$ boundary term was in fact inspired by the background subtraction boundary term for a Minkowski background.  The two boundary terms agree whenever the background subtraction boundary term is well defined, though this is a rare event for $d \ge 4$.  See \cite{MM} for a discussion of these issues.}.

Let us begin by considering a more familiar action for gravity, the Einstein-Hilbert action with Gibbons-Hawking boundary term.
\begin{equation}
\label{Szero} S_0 = \frac{1}{16\pi G} \int_M \sqrt{-g}R +
\frac{1}{8\pi G} \int_{\partial M} \sqrt{-h} K.
\end{equation}
The space + time reduction of this action is familiar, and we refer the reader to \cite{HH}
and \cite{Wald} for details. The result may be written
\begin{equation}
\label{almostADM} S_0 = \int \left( \tilde \pi^{ab} \dot{q}_{ab} -
N \tilde {\cal H} - N^a \tilde {\cal H}_a \right) + \frac{1}{8 \pi
G} \int_{\partial M} (N k - N^a \tilde \pi_{ab} r^b),
\end{equation}
where $r^b$ is the unit normal to the boundary at infinity,
$\tilde \pi^{ab}$ is the densitized momentum conjugate to the
spatial metric $q_{ab} = g_{ab} + n_a n_b$, and
\begin{equation}
\tilde {\cal H} = - \frac{ \sqrt{q}}{16 \pi G} R_\Sigma +  \frac{16
\pi G}{\sqrt q} (\tilde \pi^{ab} \tilde \pi_{ab}  -
\frac{1}{(d-2)}\tilde \pi^2),
\end{equation}
and
\begin{equation}
\tilde {\cal H}_a = - 2  D_b \tilde \pi^{ab},
\end{equation}
where $D_b$ is the covariant derivative on $\Sigma$ compatible
with $q_{ab}$. Here, following \cite{HH}, we have used a foliation
of the spacetime by surfaces $\Sigma_t$ which intersect $\partial
{\cal M}$ orthogonally.

To complete the space + time reduction of the full action (\ref{covaction}), we need now only address the final boundary term involving $\hat K$. To compute this boundary term, consider the defining
equation (\ref{Khat}):
\begin{equation}
\label{Khat2} {\cal R}_{ij} = \hat K_{ij} \hat K - \hat K_i^m \hat
K_{mj},
\end{equation}
where ${\cal R}_{ij}$ is the Ricci tensor of the boundary metric
$h_{ij}$ on $\partial M$ and where we follow the conventions of
Wald \cite{Wald}.  Note the similarity between (\ref{Khat}) and the Gauss-Codazzi
equation
\begin{equation}
\label{CGC} {\cal R}_{ij} = R_{ikjl}h^{kl} + K K_{ij} - K_{jk}
K^k{}_{i},
\end{equation}
satisfied by the extrinsic curvature $K_{ij}$.  The only
difference is the term $R_{ikjl}h^{kl}$  in (\ref{CGC}), where
$R_{ikjl}$ is the pull-back of the bulk Riemann tensor to
$\partial {\cal M}$.  As a result, it  is reasonable to compare
(\ref{Khat}) with (\ref{CGC})  and to compute the difference
$K_{ij}-\hat{K}_{ij}$ as an expansion in powers of $r$.

To lowest order in $1/r$, the Ricci curvature ${\cal R}_{ij}$ of
$\partial M$ is just that of the standard cylinder of radius $r$
in Minkowski space.  We use the coordinates indicated in
(\ref{hcyl}), so that $\mu_{ij}$ is of order $r^2$ and $n^i$ is of
order 1.  As a result,  we have \be \label{leadingKhat1}  \hat
K_{ij} = \frac{1}{r} \mu_{ij}  + O(r^{-(d-4)}). \ee It follows
from (\ref{AFdef}),(\ref{hcyl}) that the Riemann tensor in
asymptotically Cartesian coordinates is of order $r^{-(d-1)}$, so
that we also have $K_{ij} = \frac{1}{r} \mu_{ij}  +
O(r^{-(d-4)})$. These expressions define the background values
about which we wish to expand (\ref{Khat}) and (\ref{CGC}).

Since $\sqrt{-h} \sim r^{d-2}$ and  $\hat K_{ij}$ enters the
action through $\int \sqrt{-h} \hat K$, the $O(r^{-(d-4)})$ term
in $\hat K_{ij}$ also contributes to the action. However, the
$O(r^{-(d-4)})$ term is the highest order that contributes.  We
may compute this term perturbatively by linearizing
(\ref{Khatdecomp}) about the leading order term
(\ref{leadingKhat1}).

It will be useful
to decompose both $\hat K_{ij}$ and $K_{ij}$ into parts associated with the surface
$\Sigma_t$ and parts associated with the normal directions. Let us
therefore define
\begin{equation}
\label{mudef} \mu_{ij} = h_{ij} + n_i n_j,
\end{equation}
so that $\mu^i_j$ is the projector from $\partial M$ to
$\partial M \cap \Sigma_t$.  We then define
\begin{eqnarray}
\label{kdecomp}
 \hat k_{ij} = \mu_i{}^k \mu_j{}^l \hat K_{kl}, \qquad \hat M^i = \mu^{ij} n^k \hat K_{jk}, \qquad \hat M = n^j n^k
\hat K_{jk},
\cr
 k_{ij} = \mu_i{}^k \mu_j{}^l  K_{kl}, \qquad  M^i = \mu^{ij} n^k  K_{jk}, \qquad  M = n^j n^k
 K_{jk}.
\end{eqnarray}
so that we have
\begin{equation}
\hat K = \hat k - \hat M \qquad {\rm and} \qquad
 K = k - M,
\end{equation}
where $\hat k = \mu^{ij} \hat k_{ij}$, $k = \mu^{ij} k_{ij}$. From
(\ref{leadingKhat1}), we obtain \bea \label{leadingKhat} \hat
k_{ij} = \frac{1}{r} \mu_{ij}  + O(r^{-(d-4)}), & \qquad&  \hat
M_{i} = O(r^{-(d-3)}), \cr  \qquad {\rm and} \qquad  &\hat M =
O(r^{-(d-2)}) . & \eea   Again, the same relations hold to this
order for $k_{ij}, M^i, M$.  Let us denote the terms explicitly
displayed in (\ref{leadingKhat1}), (\ref{leadingKhat}) by
$K^0_{ij}, k^0_{ij}, etc$.

We decompose the boundary Ricci tensor ${\cal R}_{ij}$ similarly
as follows:
\begin{equation}
\label{rdecomp}
 \rho_{ij} := \mu_i{}^k \mu_j{}^l{\cal R}_{kl}, \qquad  {\cal S}^i := \mu^{ij} n^k {\cal R}_{jk}, \qquad {\rm and} \qquad {\cal S} :=
n^j n^k {\cal R}_{jk}.
\end{equation}
The defining equations (\ref{Khat}) and (\ref{CGC})  now yield
\begin{eqnarray}
\label{Khatdecomp}  &\rho_{ij} = (\hat k - \hat M) \hat k_{ij} -
\hat k_i{}^m \hat k_{mj} + \hat M_i \hat M_j,
\qquad
\rho_{ij} =  \mu_i^m \mu_j^n R_{mknl}h^{kl} + (
k -  M)  k_{ij} -  k_i{}^m  k_{mj} +  M_i  M_j, &\cr
& {\cal S}_i =
 \hat k \hat M_i + \hat k_{im} \hat M^m, \qquad
 {\cal S}_i =
\mu_i^m  R_{mkjl}h^{kl} n^j + k \hat M_i -  k_{im}  M^m , &\cr
&  {\cal S} = \hat k   \hat M - \hat M_m \hat M^m, \qquad
  {\cal S} = R_{ikjl}h^{kl} n^in^j +  k   M -
M_m  M^m.&
\end{eqnarray}

We note immediately that
\begin{equation}
\label{Meq} \hat M = \frac{r{\cal S}}{(d-2)} + O(r^{-(d-1)}).
\end{equation}
Furthermore, we need only the leading behavior of $n^i$ to compute
$\hat M$ to this order.  As a result,  one may treat $n^i$ as
being covariantly constant in (\ref{Meq}).  Combining this
observation with the fact that (see eqn (7.5.14) of \cite{Wald})
the linearized Ricci tensor takes the form
\begin{equation}
\label{linRicci} \delta {\cal R}_{ij} = - \frac{1}{2} h^{kl}\underline D_i \underline D_j
\delta h_{kl} - \frac{1}{2} h^{kl} \underline D_k \underline D_l \delta h_{ij} + h^{kl}
\underline D_k \underline D_{(i} \delta h_{j)l},
\end{equation}
where $\underline D_i$ is the (torsion-free) covariant derivative on
$\partial M$ compatible with $h_{ij}$, one may write $\hat M$ as a
total divergence (in $\partial M$) plus a term of order
$r^{-(d-1)}$.  As a result, $\hat M$ does not contribute to the
action $S$.  Futhermore, the only difference between $M$ and
$\hat{M}$ is
 the term $R_{ikjl}h^{kl} n^in^j$, which
is of order  $r^{-(d-1)}$. Consequently  we can use
$\Delta=k-\hat{k}$ to compute the action.

Thus, we need only calculate $\hat k$.  Clearly, since we need to solve (\ref{Khatdecomp}) to linear order, one achieves a significant simplification by focussing on  the difference $\Delta$.  One proceeds by subtracting pairs of equations in (\ref{Khatdecomp}) and obtains
\begin{equation}
\label{tildeDeltaEq2}  - \left[\mu^{ij} + n^i n^j \right]
R_{ikjl}h^{kl} = \frac{(2d-6)}{r} \Delta
\end{equation}
to leading order.

Recall that our spacetime is asymptotically flat and satisfies the vacuum Einstein equations to leading non-trivial order; i.e., the Ricci tensor vanishes at the order relevant to (\ref{tildeDeltaEq2}).
As a result, we may make the
replacement
\begin{equation}
R_{ikjl}h^{kl} \rightarrow - R_{ikjl}r^k r^l
\end{equation}
in (\ref{tildeDeltaEq2})  so that we deal only with the ``electric
part'' of the Riemann tensor (or, in fact, the Weyl tensor).  We
then use a similar argument (together with anti-symmetry of
$R_{ijkl}$) to make the further replacement
\begin{equation}
R_{ikjl} r^k r^l n^i n^j \rightarrow  R_{ikjl} r^k r^l \mu^{ij},
\end{equation}
where this time there is no change of sign because $n^i$ is
timelike.  Thus, we have
\begin{equation}
\label{tildeDeltaSol2} \Delta = \frac{2r}{2d-6}R_{ikjl} r^k
r^l \mu^{ij}.
\end{equation}

For $d=4$, precisely the expression (\ref{tildeDeltaSol2}) was considered by \cite{AM2} in comparing the Ashtekar-Hansen definition of energy to that of ADM.  They showed that, to leading order,
\begin{equation}
\label{tildeDeltaSol5} \Delta  = - \frac{1}{2} {\cal E}_{ADM} =   \frac{1}{2}
 \left( q^{ij} r^k D_k q_{ij} - q^{ik} r^j D_k
q_{ij}\right).
\end{equation}
In appendix \ref{ADM.appendix}, we give the details of this argument and show that the result (\ref{tildeDeltaSol5}) also holds for $d > 4$.

Putting all of this together with our previous results, we find
\begin{equation}
\label{ADM} S = \int \left( \tilde \pi^{ab} \dot{q}_{ab} - N
\tilde {\cal H} - N^a \tilde {\cal H}_a \right) + \frac{1}{16 \pi
G} \int_{\partial M}  \left(N \left( q^{ab} r^c D_c q_{ab} -
q^{ac} r^b D_c q_{ab}\right) - 2N^a \tilde \pi_{ab} r^b \right).
\end{equation}
This is precisely the ADM form of the
gravitational action.
 The boundary terms are just
$-N{\cal E}^{ADM} - N^a {\cal P}^{ADM}_a$.  Thus, as discussed in section
\ref{bst}, we see that the generators of asymptotic Poincar\'e transformations given in \cite{MM} are explicitly equal to the ADM generators \cite{ADM1}.


\setcounter{equation}{0}
\section{Equivalence of $\hat K$ Counterterm and the Ashtekar-Hansen Covariant Approach for $d=4$}
\label{covariant}

This section explicitly demonstrates the equality of the Ashtekar-Hansen definitions of the conserved charges with that of the boundary stress tensor defined by (\ref{covaction}) for the case $d=4$.  We expect corresponding results in higher dimensions.

\subsection{Preliminaries}

To motivate this study, recall the boundary stress tensor definition of charge (\ref{Qhyp}):
\begin{equation}
\label{Qbst}
Q[\xi] =
 \int_{C}  \sqrt{h_{C}}  T_{ij} \xi^i n^j.
\end{equation}
Applying the basic definition (\ref{Texp}) of $T_{ij}$, one
may show \cite{MM} that
\begin{equation}
\label{Tpi} T_{ij} = \frac{1}{8 \pi G} (\pi_{ij} -
\hat{\pi}_{ij}),
\end{equation}
where $\pi_{ij} = K_{ij} - K h_{ij}$ and $\hat \pi_{ij} = \hat
K_{ij} - \hat K h_{ij}$. As in section \ref{section.ADM}, one may
readily compute $\pi_{ij}- \hat \pi_{ij}$ by comparing the
Gauss-Codazzi equation for $K_{ij}$ with the defining equation
(\ref{Khat}) for $\hat K$ and expanding in powers of $\rho$.
However, in contrast to section \ref{section.ADM}, we focus here
on the case where $\partial {\cal M}$ is defined by the hyperbolic
cut-offs (\ref{hypcut}) in order to make contact with the
Ashtekar-Hansen framework\footnote{The Ashtekar-Hansen definitions
were originally stated in terms of the Spi framework
\cite{AshtekarHansen}, in which the spacetime is conformally
compactified and spacelike infinity $i^0$ is represented by a
point.  However, interesting tensor fields are not smooth at this
point.  Instead they admit direction-dependent limits.  As a
result, the fields are naturally defined on the hyperboloid of
directions in which one can approach $i^0$.  As described in
\cite{AR}, the formalism can be recast in terms in which spacelike
infinity itself is replaced by a timelike hyperboloid.  Our use of
a hyperbolic cut-off is essentially a coordinate-based description
of the same formalism.  For the convenience of the reader, we
simply translate all formulae from
\cite{AshtekarHansen,AshtekarInf, AR} into the coordinate-based
language already introduced above.}.  Nonetheless, for the same
reasons as noted in section \ref{section.ADM}, one finds \cite{MM}
that the resulting expansion for $T_{ij}$ is determined by the
expansion of the electric part of the Weyl tensor.  In particular,
\begin{equation}
\label{TandE2} T_{ij} = \frac{1}{8\pi G} \frac{\rho^{-(d-4)}
}{d-3}E_{ij} + {\cal O} (\rho^{-(d-3)}),
\end{equation}
where $E_{ac} :=  \rho^{d-3} C_{abcd} \rho^b \rho^d$ is the first non-trivial term in the electric part of the Weyl tensor. Here $\rho^a$ is the unit normal to the hyperboloid at constant $\rho$
and we have used the fact that the vacuum Einstein equations hold to leading non-trivial order.
As a result, for such $\xi$ one may rewrite (\ref{Qhyp}) as
\begin{equation}
\label{QAH2} Q[\xi] = \frac{1}{8\pi G (d-3)}
 \int_{C}  \sqrt{ h^0_{C}}  E_{ij} (\xi^0)^i  (n^0)^j,
\end{equation}
where $h^0_C$ is the determinant of the metric induced on the cut $C$ by metric on the {\it unit} hyperboloid ($ h^0_{ij} = \rho^{-2} h_{ij}$), $(n^0)^j = \rho n^j$ is the unit normal to $C$ with respect to $h^0_{ij}$, and we have introduced the corresponding $(\xi^0)^i = \rho \xi^i$.  All of the factors in (\ref{QAH2}) are now normalized so that they are independent of $\rho$ at large $\rho$ (and (\ref{QAH2}) is manifestly finite).

For $d=4$, the expression (\ref{QAH2}) is exactly the Ashtekar-Hansen definition of conserved charges corresponding to the translations.  However, less satisfactory results were obtained in \cite{MM} for the generators of Lorentz transformations.  The point is that the Ashtekar-Hansen definition of Lorentz generators involves the {\it magnetic} part of the Weyl tensor:
\begin{equation} \label{mab}
Q_{AH}[\xi] :=
 \frac{1}{8\pi G} \int_C   \sqrt{h^0_{C}} {\beta}_{ij}  \zeta^i (n^0)^j,
\end{equation}
where we have defined $\beta_{ij}$ as the pull-back to the
hyperboloid of
\begin{equation}
\label{hatbeta} {\beta}_{ab} :=  \rho^2  \epsilon^{ef}{}_{ac} C_{efbd}   \rho^c \rho^d,
\end{equation}
and the vector field
\begin{equation}
\label{zeta}
 \zeta^a := \frac{1}{2} \epsilon^{abcd}F_{cd} \rho_b
 \end{equation}
is built from the asymptotically constant  skew tensor $F_{ab}$
satisfying $\xi^a = F^{ab} \rho_b$ to leading order. The definition above is for $d=4$, though it is readily generalized to higher dimensions.
While an abstract argument for the agreement of the Lorentz charges (\ref{Qhyp}) with the usual charges was given in \cite{MM}, and while we saw explicitly in section \ref{section.ADM} that for cylindrical boundaries the charges (\ref{Qhyp}) agree with those of ADM (which in turn are known to agree \cite{AM2} with the Ashtekar-Hansen charges\footnote{Charges corresponding to boost generators were not considered in \cite{AM2}.  It is unclear to us whether this gap in the literature has been filled.}), it is far from clear precisely how (\ref{Qhyp}) and (\ref{mab}) define the same quantity.  In particular, as noted above, (\ref{Qhyp}) is fundamentally constructed from the {\it electric} part of the Weyl tensor while (\ref{mab}) is constructed from the {\it magnetic} part of the Weyl tensor.

The goal of the present section is to explicitly demonstrate the
required agreement whenever (\ref{mab}) is well-defined.  The
result will follow from relations between the electric and
magnetic parts of the Weyl tensor at appropriate orders which in
turn follow from the equations of motion.  Before commencing the
main calculation, we point out that a general metric satisfying
our definition of asymptotic flatness may have a divergent
(\ref{mab}).  This happens when, in asymptotically Cartesian
coordinates, the ${\cal O}(\rho^{-3})$ magnetic part of the Weyl
tensor is non-zero, so that $\beta_{ij}$ (in our hyperbolic
coordinates) grows with $\rho$.  For example, this occurs whenever
the spacetime carries non-zero NUT charge.  Ashtekar and Hansen
\cite{AshtekarHansen} introduced their definition only for
spacetimes in which this leading part of the magnetic Weyl tensor
vanishes, so that (\ref{mab}) is finite.  Furthermore, in such
cases by acting with an appropriate supertranslation one may
\cite{BS} impose the relation:
\begin{equation}
\label{stg}
h^1_{ij} = -2 \s h^0_{ij}.
\end{equation}
Equation (\ref{stg}) was also assumed in making the definition (\ref{mab}) \cite{AshtekarHansen}.  We therefore consider only metrics satisfying (\ref{stg})  below.  Finally, we follow \cite{AshtekarHansen} in assuming that the vacuum Einstein equations hold to order $\rho^{-4}$ in asymptotically Cartesian Coordinates.

\subsection{Asymptotic Expansions}

\label{section.old}

Because the Killing fields corresponding to asymptotic Lorentz transformations are larger at infinity than the asymptotic translations, the corresponding conserved quantities defined by (\ref{Qbst}) depend on both the leading and the next-to-leading parts of the boundary stress tensor.  Our task is to calculate these terms, and to show that the result implies agreement between (\ref{Qbst}) and (\ref{mab}).  Our calculations follow the approach used in the  systematic analysis of asymptotic
flatness was carried out  by  Beig and Schmidt in \cite{BS,
B}.   In this subsection we present the relevant asymptotic expansions for use in showing equality of the charges in sections \ref{spatial} and \ref{boosts}.

In performing the remaining asymptotic expansions, we follow \cite{BS,B} in imposing further gauge conditions to bring the metric into the form
\begin{eqnarray}
\label{metric1}ds^2 &=& N^2 d\rho^2 +  h_{ij}d\eta^i d\eta^j
\\ &=&\label{metric2} \left( 1 + \frac{\s}{\rho} \right)^2 d\rho^2 + \rho^2 \left[
h^0_{ij} + \frac{h^1_{ij}}{\rho} +  \frac{1}{\rho^2}h^2_{ij}
+ {\cal O}\left( \frac{1}{\rho^{N+1}}\right)  \right]d\eta^i d\eta^j,
\end{eqnarray} where again
$h^0_{ij}$ is the metric on the unit hyperboloid and we assume (\ref{stg}).
As stated above, we assume $d=4$ here and below.

We will make much use of the vacuum Einstein equations below.
Given the form (\ref{metric1}), it is natural to decompose these equations using the unit (outward-pointing) normal $\rho^a$ to the hyperboloid of constant $\rho$ and the projector $h_{ab} = g_{ab} -\rho_a \rho_b$.  The results may be written
\cite{BS} in the form \begin{eqnarray}
\label{fieldeq}H &:=& R +  ( K_{ij} K^{ij} - K^2) = 0, \cr
F_a &:=&\underline{D}_j (K_i^j - K \delta_i^j) = 0, \cr
F_{ij} &:=& K_{ij}{}' - 2 N \rho^2 K _{ik} K^k_j +
\underline{D}_i \underline{D}_j N - N  R_{ij} +  N K K_{ij} =0,
\end{eqnarray} where prime denotes partial derivative with respect
to $\rho$ and, as in section \ref{section.ADM}, $\underline{D}$ is the
covariant derivative compatible with the full metric $h_{ij}$ on
the hyperboloid.  In  equations (\ref{fieldeq})
indices are raised and lowered with $h_{ij}$.  However,  in the remainder of section \ref{covariant}
 indices will be raised and lowered with $h^0_{ij}$ unless
otherwise stated.

Clearly, one wishes to insert the expansion
(\ref{metric2}) into (\ref{fieldeq}) and to consider the resulting expansion of the equations of motion.
Beig \cite{B} showed that the
zeroth and the first order Einstein equations are identically satisfied
if
\begin{equation}
D^2 \s + 3 \s =0. \label{sigma}
\end{equation}
Here we have introduced the (torsion-free) covariant derivative $D_i$ on the hyperboloid compatible with the zero-order metric $h^0_{ij}$.
Turning to the second order equations, \cite{B} showed that these may be written in the form
\begin{eqnarray}
\label{beig20a}h^2_i{}^i &=& 12 \s^2 + \s_i \s^i, \\
\label{beig20b} D_j h^2_i{}^j &=& 16 \s \s_i + 2 \s_j \s^j_i, \\
\label{beig20c}D^2 h^2_{ij} - 2 h^2_{ij} &=& 6 \s_k \s^k h^0_{ij} +
8 \s_i \s_j + 14 \s \s_{ij} - 18 \s^2 h^0_{ij} + 2 \s_{ik} \s^k_j +
2 \s_{ijk} \s^k,
\end{eqnarray}
where $\s_i = D_i \sigma$, $\s_{ij} = D_j D_i \sigma$, and
$\s_{ijk} = D_k D_j D_i \s$.

The expansion (\ref{metric1}) allows us to write the electric and magnetic parts of the Weyl tensor in the form
\begin{eqnarray}
E_{ij} &=& - \sigma_{ij} - \sigma h^0_{ij} + {\cal O}(\rho^{-1}), \label{elec} \cr
\beta_{ij} &=&  \epsilon_{mni} D^m\left(h^2_j{}^n - 2 \s^2
\delta_j^n\right)  + {\cal O}(\rho^{-1}). \label{beta}
\end{eqnarray}
where it is easy to show that $\beta_{ij} =\beta_{ji}$ using
(\ref{beig20b}), (\ref{beig20c}).  Here, $\epsilon_{mni}$ is the
Levi-Civita tensor on the unit hyperboloid (metric $h^0_{ij}$).
The first equation is straightforward to derive.  The second was
used in \cite{B} and is re-derived for completeness in appendix
\ref{pi} and \ref{sectionidentity}.  A useful fact is that the
equations of motion (\ref{beig20a}-\ref{beig20b}) imply that
expression (\ref{beta}) for $\beta_{ij}$ is symmetric in $i,j$.

One may make use of (\ref{beta}) to write
(\ref{beig20a}-\ref{beig20c}) in the form
\begin{eqnarray} \label{beig22a}(D^2 - 2)\beta_{ij} &=& -
4 \epsilon_{kl(i}
\s^k E_{j)}^l \\ \label{beig22b} \beta_i^i & =& 0 \\
\label{beig22c}D_j \beta_i^j &=& 0. \end{eqnarray}
The
systems of equations (\ref{beig20a}-\ref{beig20c}) and
(\ref{beig22a}-\ref{beig22c}) are equivalent \cite{B} and the explicit
transformation from the second to the first is via the following
change of variables
\begin{equation}
\label{curlb} h^2_{ij} =  - \epsilon_{kli} D^k \beta_{j}{}^l + 6
\s^2 h^0_{ij} - 2 \s_i \s_j + \s_k \s^k h^0_{ij} + 2 \s \s_{ij}.
\end{equation}
This transformation will be useful in sections \ref{spatial} and \ref{boosts}.

\subsection{Spatial Rotations}

\label{spatial}

With the aid of the asymptotic expansions of section \ref{section.old}, we now proceed to show the equivalence of the two
covariant approaches for the remaining charges. In this subsection we concentrate on spatial rotations
and in the next subsection we discuss the case of the boost
generators. The arguments in both of these subsections are similar
in spirit but they differ significantly in the details.

Recall that the definition (\ref{mab}) of $Q_{AH}[\xi]$ is given
in terms of a ``dual Killing field'' $\zeta$, defined by
(\ref{zeta}).  As a result, before beginning the main calculation
it will be useful to follow \cite{AM2} in presenting an alternate
form of the relation between $\xi^a$ and $\zeta^a$ adapted to the
case where $\xi^a$ is a spatial rotation. Let $\xi^i$ be a
rotational Killing field on the unit hyperboloid $(h_{ij}^0, {\cal
H}$). A time coordinate $t$ is naturally induced on this
hyperboloid by its embedding in Minkowski space.  Consider the
cross section $C_0$ at $t=0$, and let the unit (future-pointing)
normal to $C_0$ in ${\cal H}$ be $(n^0)^i$. Note that  $\xi^i$ is
also a Killing field on the cut $C_0$, which is just $S^2$ with
the round metric. A Killing field on $S^2$ with round metric can
be written in terms of the derivatives of a function $f$ on $S^2$
as
\begin{equation}
\label{Killing} \xi^i = \epsilon^{ij}_2 D^{(S^2)}_j f,
\end{equation}
where $D^{(S^2)}_j$ and $\epsilon^{ij}_2$ are the covariant derivative and Levi-Civita tensor on the round $S^2$.
The boost Killing field $\zeta^i$ on $(h_{ij}^0, {\cal H})$ defined by (\ref{zeta}) then satisfies
\be \zeta^i  = f (n^0)^i \qquad {\rm at} \ t=0
\label{Killing2}. \ee We will see that the appearance of the dual boost Killing field in the
Ashtekar-Hansen definition of angular momentum can traced to the Hodge star in the
definition of the magnetic part of the Weyl tensor (\ref{hatbeta}).

We now proceed with the main calculation, taking the cut $C$ to be
$C_0$ above. Inserting the relation (\ref{Tpi}) into (\ref{Qbst}),
one may write the counter-term charge in the form
\begin{equation}
\label{firstdp}
 Q[\xi] = \frac{1}{8\pi G}  \int_{C} \Delta \pi_{ij} \xi^i  n^j  \sqrt{h_C}
= \int_{C} \left( \frac{\Delta \pi_{ij}^1}{\rho} + \frac{\Delta \pi_{ij}^2}{\rho^2} \right) \xi^i (n^0)^j \rho^2 \left( 1 +
\frac{\s}{\rho} \right) \left( 1- \frac{2 \s}{\rho} \right)
\sqrt{h_C^0},
\end{equation}
where $\Delta \pi_{ij} = \pi_{ij} - \hat \pi_{ij}$  and we have
introduced the expansion $\Delta \pi_{ij}  = \sum_{n > 0} \Delta
\pi^n_{ij} \rho^{-n}$.  In deriving (\ref{firstdp}),   we have
used (\ref{stg}) and the fact that $\Delta \pi_{ij}^0=0$; we have
also dropped terms which do not contribute in the limit $\rho \to
\infty$.

The integrand of (\ref{firstdp}) is of order $\rho$.  However, we have seen that $\Delta \pi^1_{ij}$ is given by the electric part of the Weyl tensor.  As a result, one may use (\ref{elec}) to show that this term gives no contribution to the integral \cite{AshtekarHansen,AshtekarInf,AR}.  Thus we may write
\begin{equation}
\label{Qdpt}
 Q[\xi] = \int_C \Delta \tilde \pi^2_{ij} \xi^i
(n^0)^j \sqrt{h_C^0}
 + {\cal O}(\rho^{-1}),
\end{equation}
 where
\be
\Delta \tilde \pi^2_{ij} = \Delta \pi^2_{ij} - \s
\Delta \pi^1_{ij}. \ee
It is shown in appendix
\ref{pi} that $\Delta \tilde \pi^2_{ij}$ is divergence free with
respect to derivative $D^i$ so that (\ref{Qdpt}) is independent of the choice of cut $C$.

Now, when $\xi^i$ is a rotation we may insert expression
(\ref{Killing}) into (\ref{Qdpt}) and perform an integration by
parts to find:
 \begin{equation}
\label{above1}
 Q  [\xi] = - \frac{1}{8\pi G}  \int_C \epsilon_2^{mn} D_{m} \Delta \tilde \pi^2 _{nj}  f  (n^0)^j \sqrt{h_C^0}.
\end{equation}
Here we have used the fact that $D_m^{(S^2)} (n^0)^j=0$.  We now express $\epsilon_2^{mn}$ in (\ref{above1}) in terms of the Levi-Civita tensor $\epsilon^{mnk}$ on the unit hyperboloid:
\be \label{above2} Q [\xi] = -  \frac{1}{8\pi G}\int_C
\epsilon^{mn}{}_{(i} D_{|m}\Delta \tilde \pi^2_{n|j)} f  (n^0)^i (n^0)^j\sqrt{h_C^0}  = - \frac{1}{8\pi G}
\oint_C \epsilon^{mn}{}_{(i} D_{|m}\Delta \tilde \pi^2_{n|j)}
\zeta^i (n^0)^j  \sqrt{h_C^0}, \ee
where in the second step we have used
expression (\ref{Killing2}) for the Killing field $\zeta^i$ .

From (\ref{above2}), the next step is to carefully expand $\Delta \tilde \pi^2_{ij}$ and to use (\ref{beta}) and (\ref{curlb}) to express the results in terms of $\beta_{ij}$.  This is somewhat tedious, and we have relegated such calculations to appendix \ref{pi}.  For our present purposes, the key result is the  identity (\ref{rotation}):
\begin{eqnarray}
{\epsilon}^{mn}{}_{(i}{D}_{|m}\Delta \tilde{\pi}^2_{n|j)} &=&
 - \beta_{ij} -\frac{1}{2}(D^2 + 2) \left( \epsilon^{mn}{}_{(i}
\s_{|n} \s_{m|j)}\right). \label{rotation1}
\end{eqnarray}
Finally, using
 (\ref{beig22a}) we note that
\begin{equation}
4 \epsilon^{mn}{}_{(i} \s_{|n} \s_{m|j)} \zeta^i    = \zeta ^{i}   (D^{2}-2)\beta _{ij} =2D^{i}\left( \zeta ^{k}D_{[i}\beta
_{j]k}+\beta _{k[i}D_{j]}\zeta ^{k}\right) \label{Beig30}
\end{equation}%
and
\begin{equation}
4 \epsilon ^{mn}{}_{(i}D^{2}(\sigma _{|n}\sigma _{m|j)})\zeta
^{i}=  \zeta ^{i} D^{2}((D^{2}-2)\beta _{ij}) =2D^{i}\left( \zeta
^{k}D_{[i}D^{2}\beta _{j]k}+D^{2}\beta _{k[i}D_{j]}\zeta ^{k}\right).
\label{Beig30p}
\end{equation}%
Thus, the second term in (\ref{rotation1}) contributes only a total divergence on $C$ to the integrand of (\ref{above2}).  We have derived
\begin{equation}
\label{2terms}
 Q  [\xi] =   \frac{1}{8 \pi G}  \int_{C} \beta_{ij} \zeta^i t^j \sqrt{h_C^0} = Q_{AH}[\xi],
\end{equation}
as desired.

\subsection{Boosts}
\label{boosts} Let us now consider the case where $\xi$ is a
boost. The argument for equality of $Q[\xi]$ and $Q_{AH}[\xi]$ is
similar in spirit to that given for rotations above, though the
details are somewhat different.  Again, since the definition
(\ref{mab}) of $Q_{AH}[\xi]$ makes use of the dual Killing field
$\zeta$, we begin by rewriting the relationship between $\xi$ and
$\zeta$. Since $\xi$ is a boost, on the cross section $C_0$ of the
hyperboloid   $\xi^i$ is
 proportional to $(n^0)^i$. Set $\xi^i = g (n^0)^i$ where $g$ is a
function on $S^2$. Then the dual Killing field is the rotation
$\zeta^i = \epsilon^{in}_2 D_n g$.  Again, we take $C$ to be $C_0$
below.

We now rewrite the Ashtekar-Hansen charge as
\begin{equation}
Q_{AH}[\xi] =   \frac{1}{8 \pi G}  \int_C \epsilon^{jn}{}_k
\beta_{ij} (D_n g) (n^0)^i (n^0)^k \sqrt{h_C^0}.
\end{equation}
Integrating by parts yields
\begin{eqnarray}
Q_{AH}[\xi] &=& -  \frac{1}{8 \pi G} \int_C \epsilon^{jn}{}_{(k}
D_{|n} \beta_{j|i)} g (n^0)^i (n^0)^j \sqrt{h_C^0}
\\ &=&  \frac{1}{8 \pi G}  \int_C \epsilon^{mn}{}_{(i} D_{|m} \beta_{n|j)} \xi^i (n^0)^k \sqrt{h_C^0} \label{460}
\end{eqnarray}
where in the last step we have used the expression $\xi^i = g (n^0)^i$.

Again a careful expansion of $\pi_{ij}, \hat \pi_{ij}$ is required, for which we refer the reader to appendix \ref{pi}.  This time the key result is the relation (\ref{deltatildepi2}):
\begin{equation}
\label{deltatildepi3} \Delta \tilde \pi^2_{ij}  = \epsilon_{kl(i}
D^k \beta_{j)}{}^l - 2 \s \s_{ij} - \frac{5}{2} \s^2 h^0_{ij} +
\frac{1}{2} h^0_{ij} \s_{kl} \s^{kl} - \s_{a}{}^r \s_{rj}.
\end{equation}
Applying this to (\ref{460}) yields
\begin{equation}
 Q_{AH}[\xi] = \frac{1}{8\pi G}
 \int_C \left(\Delta \tilde \pi^2_{ij } + 2 \s \s_{ij} +
\frac{5}{2} \s^2 h^0_{ij} + \s_i^k \s_{kj} - \frac{1}{2} \s_{kl}
\s^{kl} h^0_{ij} \right) \xi^i (n^0)^j  \sqrt{h_C^0}. \label{old}
\end{equation}
 A bit of calculation shows that the unwanted terms in (\ref{old}) satisfy
\begin{equation}
\left(2 \s \s_{ij} + \frac{5}{2} \s^2 h^0_{ij} + \s_i^k \s_{kj} -
\frac{1}{2} \s_{kl} \s^{kl} h^0_{ij} \right) \xi^i =  D^i \left(
\xi^k D_{[i} k_{j]k} + k_{k[i} D_{j]} \xi^k \right) +  j_{ij}\xi^i , \
\end{equation}
 where
 \begin{equation}
 k_{ij} := - \frac{1}{2} \s^k
\s_k h^0_{ij} + \s_i \s_j + \frac{3}{2} \s^2 h^0_{ij} \qquad {\rm and}
 \end{equation}
\be
  j_{ij} := 2 (\s \s_{ij} + 2 \s^2 h^0_{ij} + \s_i \s_j - \s_k
\s^k h^0_{ij}).
 \end{equation}
 Note that  both $k_{ij}$ and $j_{ij}$
are divergence free.  Furthermore, one may verify the relation
\begin{equation}
 j_{ij}\xi^i  = D^i ( \s^2 D_{[i} \xi_{j]})  - 4 D^i ( \s
\s_{[i} \xi_{j]}).
 \end{equation}
 Thus, the integrand of (\ref{old}) is just $\Delta \tilde  \pi_{ij} \xi^i(n^0)^j$ plus a total divergence on $C$.
As desired, we have demonstrated the explicit agreement between the Ashtekar-Hansen charges and the charges given in \cite{MM}.

\section{Discussion}
\label{disc}

The above work has answered certain open questions relating to  the
variational principle (\ref{covaction}) proposed in \cite{MM} for
asymptotically flat spacetimes, and to the associated conserved
charges.  While it was argued in \cite{MM} on general grounds that
the Poincar\'e generators\footnote{There were some caveats to the
agreement for boost generators.} defined by the boundary stress
tensor should agree with those defined by other approaches
\cite{ADM1}-\cite{MM},
the agreement was shown explicitly only for energy and momentum.
Our work here also makes the agreement explicit for the
Lorentz generators.  For pedagogical reasons and as a consistency
check, we have also separately shown agreement with the canonical
(i.e., ADM \cite{ADM1}) generators and with the covariant
generators of Ashtekar and Hansen \cite{AshtekarHansen}.

In particular, we first showed for $d \ge 4$ that the Legendre
transform of (\ref{covaction}) is the ADM Hamiltonian \cite{ADM1},
with precisely the ADM boundary terms.
 We used this fact to explicitly demonstrate
that the Poincar\'e generators defined in \cite{MM} agree with
those of ADM \cite{ADM1}.  Note that the corresponding statement
does not hold in $d=3$ spacetime dimensions.  There the analogous
Hamiltonians differ by a constant which shifts the energy
of 2+1 Minkowski space \cite{DonLeo}.

Second, for $d=4$ we have answered an open question related to the
Lorentz generators.  The explicit agreement of boundary stress
tensor energy and momentum with the Ashtekar-Hansen definitions
was noted in \cite{MM}.  However, the agreement of the Lorentz
generators was more mysterious.  In \cite{MM}, these generators
were given in terms of the electric part of the Weyl tensor, while
the Ashtekar-Hansen definition \cite{AshtekarHansen} was stated in
terms of the magnetic part of the Weyl tensor.  Our work resolves
this tension by showing that the Einstein equations relate the
higher-order parts of the electric and magnetic Weyl tensors in
precisely the right way to obtain agreement.  We expect
corresponding results in higher dimensions.

\subsubsection*{Acknowledgments} We would like to thank Abhay Ashtekar
for a number of useful discussions concerning asymptotic flatness. AV would also like to thank Keith Copsey for several useful discussions.
This research was supported in part by the National Science
Foundation under Grants No. PHY99-07949, No. PHY03-54978, by funds
from the University of California and by the Natural Sciences and Engineering Research Council of Canada.

\appendix

\setcounter{equation}{0}
\section{Generalizing Ashtekar-Magnon to $d > 4$}
\label{ADM.appendix}

This appendix contains the remaining details showing the equality of the covariant action (\ref{covaction}) with the $\hat K$ counter-term and the ADM canonical action (\ref{can}).  In particular, section \ref{section.ADM} makes use of the result
\begin{equation}
\label{AM2result} \Delta  = - \frac{1}{2} {\cal E}_{ADM} =   \frac{1}{2}
 \left( q^{ij} r^k D_k q_{ij} - q^{ik} r^j D_k
q_{ij}\right),
\end{equation}
to leading order.
This result was derived in \cite{AM2} for the case $d=4$.  Below, we give the details of this argument and show that (\ref{AM2result}) also holds for $d > 4$.  All equations below are valid to leading order.

We begin with equation (\ref{tildeDeltaSol2}) from section
\ref{section.ADM}:
\begin{equation}
\label{from4} \Delta = \frac{2r}{2d-6}R_{ikjl} r^k
r^l \mu^{ij}.
\end{equation}
It is useful to rewrite (\ref{from4}) in terms of the Riemann tensor
$(R_\Sigma)_{ijkl}$ associated with the hypersurface $\Sigma_t$.
This can be done by using the Gauss-Codazzi equation for $\Sigma$ as a hypersurface in ${\cal M}$. Since in
asymptotically Cartesian coordinates the extrinsic curvature
$\nabla_i n_j$ of $\Sigma_t$ falls off as $r^{-(d-2)}$ we have
\begin{equation}
R_{ikjl} r^k r^l \mu^{ij} = (R_\Sigma)_{kl} r^k r^l +
O(r^{-2(d-2)}).
\end{equation}
We desire only the leading behavior, so we may now use equation
(\ref{linRicci}) to compute the Ricci tensor of $\Sigma$. The
result is
\begin{equation}
\label{tildeDeltaSol3} \Delta = \frac{r}{2d-6} r^k r^l
\left[ -q^{ij} D_k D_l q_{ij} - q^{ij} D_i D_j q_{kl} + 2 q^{ij}
D_i D_k q_{jl}  \right],
\end{equation}
where $D_i$ is the covariant derivative on $\Sigma_t$.

Since $\mu_{ij} = q_{ij} - r_i r_j$, we may rewrite
the above equation as
\begin{eqnarray}
\label{tildeDeltaSol4} \Delta &=& \frac{r}{2d-6} [D_k D_l
q_{ij}] \left(  -q^{ij} r^k r^l - q^{kl} r^i r^j + q^{jl} r^i r^k
+ q^{ik} r^j r^l \right) \cr &=& \frac{r}{2d-6} [D_k D_l q_{ij}]
\left( -q^{ij} r^k r^l - \mu^{kl} r^i r^j + q^{jl} r^i r^k
 + \mu^{ik} r^j r^l \right). \ \ \ \
\end{eqnarray}

Note that, when integrated over the sphere, the last term may be
written
\begin{eqnarray}
\label{term4} &\int_{S^{d-2}}& \mu^{ik} r^j r^l D_k D_l q_{ij}
\cr &=& \int_{S^{d-2}} \mu^{ik} r^jD_k \left( r^lD_l q_{ij}
\right) - \int_{S^{d-2}} \mu^{ik} r^j \left(
\frac{\delta_k^l}{r} - \frac{r_k r^l}{r} \right) D_l q_{ij} \cr
&=& \int_{S^{d-2}} \mu^{ik} r^jD_k \left( r^lD_l q_{ij} \right)
- \frac{1}{r} \int_{S^{d-2}} \mu^{ik} r^j  D_k q_{ij} \cr &=&
\int_{S^{d-2}} \mu^{ik} D_k \left(r^j  r^lD_l q_{ij} \right) -
\frac{1}{r} \int_{S^{d-2}} \mu^{ik} \left(D_k  r^j\right)  r^l
D_l q_{ij} - \frac{1}{r} \int_{S^{d-2}} \mu^{ik} r^j  D_k
q_{ij} \cr &=& - \frac{1}{r} \int_{S^{d-2}} \mu^{ik}   r^l D_l
q_{ik} - \frac{1}{r} \int_{S^{d-2}} \mu^{ik} r^j  D_k q_{ij},
\end{eqnarray}
where in the final step we have used the fact that the first term
in the 2nd to last line is a total divergence on the sphere. Thus,
when integrated over the sphere it gives zero. In earlier steps,
we used the fact that we require only the leading term in $1/r$ to
make the replacement $D_k r^l \rightarrow \left(
\frac{\delta_k^l}{r} - \frac{r_k r^l}{r} \right).$

Similarly, we can write the second term in (\ref{tildeDeltaSol4})
as
\begin{eqnarray}
\label{term2} \frac{1}{r} \int_{S^{d-2}} \mu^{kl}   r^i D_l
q_{ik} + \frac{1}{r} \int_{S^{d-2}} \mu^{kl} r^j  D_l q_{kj}.
\end{eqnarray}
Adding (\ref{term2}) and (\ref{term4}) yields:
\begin{equation}
\label{2+4}  \frac{1}{r} \int_{S^{d-2}}  \left( \mu^{jk}   r^i
D_k q_{ij} - \mu^{ij} r^k  D_k q_{ij} \right) = \frac{1}{r}
\int_{S^{d-2}}  \left( q^{jk}   r^i D_k q_{ij} - q^{ij} r^k D_k
q_{ij} \right).
\end{equation}

We now turn to the first and third terms  from
 (\ref{tildeDeltaSol4}).
Using the fact that
\begin{equation}
q_{kl} = \delta_{kl} + \frac{q^1_{kl}}{r^{(d-3)}} + O(r^{-(d-2)}),
\end{equation}
where $q^1_{kl}$ is independent of $r$, to leading order we may make the replacement
\begin{equation}
r^i  D_i D_j q_{kl} \rightarrow - \frac{d-2}{r} D_j q_{kl}
\end{equation}
in these terms.

Finally, since we require only the leading order behavior (while
any Christoffel symbol is of order $r^{-(d-2)}$), we may also
commute derivatives freely in (\ref{tildeDeltaSol4}).
We thus arrive at
\begin{equation}
\label{tildeDeltaSol52} \Delta \rightarrow \frac{r}{2d-6}
\frac{d-3}{r} \left( q^{ij} r^k D_k q_{ij} - q^{ik} r^j D_k
q_{ij}\right),
\end{equation}
which agrees with (\ref{AM2result}), as desired.


\setcounter{equation}{0}
\section{Expansions of $\pi_{ij}$, $\hat{\pi}_{ij}$ and $\Delta \pi_{ij}$}

\label{pi}

In this appendix we outline a method for calculating $\pi_{ij}$,
$\hat\pi_{ij}$ and $\Delta \pi_{ij}$. We confine ourselves to four
spacetime dimensions and we work with the Beig-Schmidt coordinate
charts discussed in section \ref{section.old}.

Let us begin with the extrinsic curvature $K_{ij}$:
\begin{equation}
K_{ij} =  \frac{\rho}{2N} \left[ \rho h'_{ij} + 2 h_{ij} \right] =
\rho \left[ 1- \frac{2\s}{\rho} + \frac{2 \s^2}{\rho^2} \right]
h_{ij}^0 + {\cal O}\left( \frac{1}{\rho^2}\right),
\end{equation}
whose trace is
\begin{equation}
K   = h^{ij}K_{ij} = \frac{3}{\rho} + \frac{1}{\rho^3} \left( 6 \s^2
+ \s_k \s^k \right) + {\cal O}\left( \frac{1}{\rho^4}\right).
\end{equation}
where (\ref{beig20a}) has been used. Next, we calculate $\pi_{ij}$:
\begin{eqnarray}
 \pi_{ij} &=& K_{ij} - h_{ij} K \\
&=& -2 h^0_{ij} \rho +  \left[ 4 \s h^0_{ij} \right] +
\frac{1}{\rho} \left[ - 4 \s^2 h^0_{ij} - 3 h^2_{ij} + h^0_{ij}
h_k^{2k} \right] + {\cal O}\left( \frac{1}{\rho^2}\right).
\label{pieqn}
\end{eqnarray}

The calculation of $\hat \pi_{ab}$ is similar, but somewhat more
involved.  It is convenient to introduce $\hat p_{ij} =
\frac{1}{\rho} \hat K_{ij}$. Recall that the counterterm $\hat K$
is defined implicitly via the Gauss-Codacci like equation (\ref{Khat2})
\begin{equation}
\label{counterterm}R_{ij}  = \hat K_{ij} \hat K - \hat K_{ik} \hat
K_{jl} h^{kl} = \hat p_{ij} \hat p_{kl} \tilde h^{kl} - \hat p_{ik}
\hat p_{jl} \tilde h^{kl}.
\end{equation}
where  $ \tilde h_{kl} := \rho^{-2} h_{kl} $ is a conformally
rescaled metric on the hyperboloid. In the remainder of this section
raising and lowering of indicies is done with $h^0_{ij}$. Expanding
$\hat p_{ij}$ as
\begin{equation}
\hat p_{ij} = h^0_{ij} + \frac{1}{\rho} \hat p^1_{ij} +
\frac{1}{\rho^2} \hat p^2_{ij} + {\cal O} \left(
\frac{1}{\rho^3}\right), \label{expansionp}
\end{equation}
we find that the right-hand side of (\ref{counterterm}) is
\begin{eqnarray}
\nonumber R_{ij} &=& 2 h^0_{ij} + \frac{1}{\rho} \left(  \hat
p^1_{ij} + (\hat p^1 - h^1) h^0_{ij} + h^1_{ij} \right) \\ & + &
\frac{1}{\rho^2} \Bigg{(} \hat p^2_{ij} + h^0_{ij} ( \hat p^2 - h^2
+ h^{1kl}h^1_{kl} ) +
h^2_{ij} - h^1_i{}^k h^1_{kj} + \hat p^1_{ij} ( \hat p^1 -h^1) \nonumber \\
&-& h^0_{ij} \hat p^{1kl} h^1_{kl} - p^1_{i}{}^{k} \hat p^1_{kj} +
h^1_{i}{}^{k} \hat p^1_{kj} +h^1_{j}{}^{k} \hat p^1_{ki} \Bigg{)} +
{\cal O}\left( \frac{1}{\rho^3} \right)
\end{eqnarray} We can also invert the above relation
to express $\hat p^1_{ij}, \hat p^2_{ij}$ in terms of the
expansion $R_{ij} = \sum_{n > 0} R_{ij}^n \rho^{-n}$ of the Ricci
tensor on the hyperboloid. Comparing the first order terms we
obtain
\begin{equation}
R^1_{ij} = \hat p^1_{ij} + h^1_{ij} + h_{ij}^0 ( \hat p^1 - h^1).
\label{r1}
\end{equation}
Taking the trace of (\ref{r1}) we can express $\hat p^1$ in terms of
$R^1$ and then we can reexpress $\hat p^1_{ab}$ in terms of
$R^1_{ab}$. We find
\begin{equation}
\hat p^1_{ij} = R^1_{ij} - \frac{1}{4} h^0_{ij} R^1 - h^1_{ij} +
\frac{1}{2} h^0_{ij} h^1 \label{hatp1}.
\end{equation}
A similar calculation for $\hat p^2_{ij}$ gives
\begin{eqnarray}
\hat p^2_{ij} &=& R^2_{ij} - \frac{1}{4} h^0_{ij} R^2 + \frac{1}{2}
h^0_{ij} ( h^2  -  h^1_{kl} h^{1kl}) - h^2_{ij}  + h^1_{il} h^{1l}_j
- \hat p^1_{ij} (\hat p^1 - h^1) + \hat p^1_{i}{}^k \hat p^1_{kj}
\cr & & -\hat p^1_{i}{}^k h^1_{kj}- \hat p^1_{j}{}^k h^1_{ki}  +
\frac{1}{4} h^0_{ij} \left(\hat p^1 ( \hat p^1 - h^1) + 3\hat
p^{1kl}h^1_{kl} -\hat p^{1kl}\hat p^1_{kl}\right) \label{hatp2}.
\end{eqnarray}
With these quantities in hand we can easily calculate $\hat
\pi_{ij}$. Using the definition of $\hat \pi_{ij}$ and the
expansion of $\hat p_{ij}$ and $\tilde h_{ij}$ we have
\begin{eqnarray}
\nonumber
\frac{1}{\rho} \hat \pi_{ij} &=& \hat p_{ij} - \tilde h_{ij} \hat p \\
&=& \nonumber -2 h^0_{ij} + \frac{1}{\rho} \left( \hat p^1_{ij} -
h^0_{ij} \hat p^1 - 3 h^1_{ij} + h^0_{ij} h^1 \right)
\\ \nonumber &+& \frac{1}{\rho^2} \left[ \hat p^2_{ij} - h^0_{ij} \hat p^2 - 3
h^2_{ij} + h^0_{ij} ( h^2 - h^{1kl}h_{kl}^1) + h^1_{ij} h^1 +
h^0_{ij} h^{1kl} \hat  p^1_{kl} - h^1_{ij} \hat p^1 \right] \\ &+&
{\cal O}\left( \frac{1}{\rho^3} \right)\label{hatpi}.
\end{eqnarray}
Substituting  (\ref{hatp1}), (\ref{hatp2}),
 and (\ref{beig20a}-\ref{beig20c}) into
(\ref{hatpi}) and making use of various identities from appendix \ref{collection} we find
\begin{equation}
\hat \pi^1_{ij} = \s_{ij} + 5 \s h^0_{ij}
\end{equation}
and
\begin{eqnarray}
\hat \pi^2_{ij}& =& - \frac{13}{2} \s^2 h^0_{ij} - 2 h^2_{ij} +
h^0_{ij} h^2 - \s_c \s^c h^0_{ij} + 2 \s_i \s_j + \s \s_{ij} +
\s_i{}^k\s_{kj} - \frac{1}{2} \s^{kl}\s_{kl} h^0_{ij}.
\end{eqnarray}
Therefore,
\begin{eqnarray}
\Delta \pi_{ij} &=& \pi_{ij} - \hat \pi_{ij} = \frac{E^0_{ij}}{\rho}
+ \frac{\Delta \pi^2_{ij}}{\rho^2} + {\cal
O}\left(\frac{1}{\rho^3} \right) \label{above5} \\
\Delta \pi^2_{ij} &=& \left( \frac{5}{2} \s^2 + \s_k \s^k +
\frac{1}{2} \s_{kl} \s^{kl} \right) h^0_{ij} - h^2_{ij} - 2\s_i
\s_j - \s \s_{ij} - \s_{i}{}^k\s_{kj}
\label{above3} \end{eqnarray} where
$E^0_{ij}$ is the leading order electric part of the Weyl tensor
\begin{equation}
E^0_{ij} = -\s_{ij} - \s h_{ij}^0.
\end{equation}
Inserting (\ref{curlb}) into (\ref{above3}) yields the useful expression
\begin{equation}
\label{above4}
\Delta \pi^2_{ij} = \epsilon_{kli} D^k \beta_{j}{}^l - 3 \s \s_{ij}
- \frac{7}{2} \s^2 h^0_{ij} + \frac{1}{2} h^0_{ij} \s_{kl} \s^{kl} -
\s_{i}{}^k \s_{kj}.
\end{equation}
From (\ref{above5}), it can be easily
checked that $ \underline{D}_i \Delta \pi^{ij} = 0$. However, $D^i
\Delta \pi^{2}_{ij} \neq 0$.

We now consider $\Delta \tilde \pi^2_{ij} $: \bea
\label{deltatildepi} \Delta
\tilde \pi^2_{ij}  &=&  \Delta \pi^2_{ij} - \s E_{ij} \\
\label{deltatildepi2} &=& \epsilon_{kli} D^k \beta_{j}{}^l - 2 \s
\s_{ij} - \frac{5}{2} \s^2 h^0_{ij} + \frac{1}{2} h^0_{ij} \s_{kl}
\s^{kl} - \s_{i}{}^k \s_{kj}. \eea It is straightforward to verify
that (\ref{deltatildepi2}) is divergence-free with respect to the
derivative $D_i$, i.e., $D^i \Delta \tilde \pi^2_{ij} = 0$.

Finally,  we note that
\begin{eqnarray}
{\epsilon}^{mn}{}_{(i}{D}_{|m}\Delta \tilde{\pi}^2_{n|j)} &=&
-{\epsilon}^{mn}{}_{(i}D_{|m}h^2_{n|j)}
-{\epsilon}^{mn}{}_{(i}\s_{j)m}\s_n
-\s_n^k{\epsilon}^{mn}{}_{(i}\s_{j)km}  \\
&=& -{\epsilon}^{mn}{}_{(i}D_{|m}h^2_{n|j)}
-\frac{1}{2}\epsilon^{mn}{}_{(i} (D^2 + 2) (\s_{|n} \s_{m|j)})
\end{eqnarray}
Using (\ref{beta}), one may verify the result
\begin{equation}
{\epsilon}^{mn}{}_{(i}{D}_{|m}\Delta \tilde{\pi}^2_{n|j)} =
- \beta_{ij} -\frac{1}{2}(D^2 + 2) \left( \epsilon^{mn}{}_{(i}
\s_{|n} \s_{m|j)}\right), \label{rotation}
\end{equation}
which is central to the argument in section \ref{spatial}.


\setcounter{equation}{0}
\section{An Identity for $\pi_{ij}$}
\label{sectionidentity} In this appendix we derive the relation
(\ref{beta}).  We start by proving an  identity (\ref{identity})
relating the magnetic part of the Weyl tensor to the curl of the
conjugate momentum $\pi_{ij}$. Then we calculate the curl of the
conjugate momentum and express it in terms of $ h^2_{ij}$ thus
establishing the relation (\ref{beta}).  In proving the identity
(\ref{identity}) we closely follow the proof of Gauss Codacci
equations as given in \cite{Wald}. Let's consider a timelike surface
with unit outward pointing normal $n^a$. Then, the extrinsic
curvature satisfies
\begin{eqnarray}
\underline{D}_m K_{nj} - \underline{D}_n K_{mj} &=&   h_m{}^p
  h_n{}^q   h_j{}^l \left( \n_p K_{ql} - \n_q K_{pl}
\right) \cr &=&   h_m{}^p   h_n{}^q   h_j{}^l \left(
\n_p \left(   h_q{}^r \n_r n_l
\right) - \n_q \left(   h_p{}^r \n_r n_l \right) \right) \nonumber \\
&=&     h_m{}^p   h_n{}^r   h_j{}^l
 R_{prls}n^s =  R_{mnjs}n^s,
\end{eqnarray}
where following the standard conventions we have used the same
symbol $R_{mnbf}$ for the projected Riemann tensor. Now consider
\begin{eqnarray}
\underline{D}_m \pi_{nj} - \underline{D}_n \pi_{mj} &=&
\underline{D}_m K_{nj} - \underline{D}_n K_{mj} - \tilde
h_{nj}\underline{D}_m K + \tilde h_{mj} \underline{D}_nK \cr &=&
R_{mnjs}n^s - \tilde h_{nj}\underline{D}_m K + \tilde h_{mj}
\underline{D}_nK.
\end{eqnarray}
Multiplying with the Levi-Civita tensor on the hyperboloid (metric
$h_{ij}$), $\underline{\epsilon}^{mn}{}_i$, and symmetrizing over
$i$ and $j$ we get
\begin{eqnarray}
\underline{ \epsilon}^{mn}{}_{(i} \underline{D}_{|m} \pi_{n|j)} =
\frac{1}{2} \underline{  \epsilon}^{mn}{}_{(i} R_{|mn|j)s}n^s =
\frac{1}{2} \underline{\epsilon}^{mn}{}_{(i} C_{|mn|j)s} n^s.
\label{iden}
\end{eqnarray}
Finally, notice that the Levi-Civita tensor on the hyperboloid is
related to the spacetime Levi-Civita tensor via
\begin{equation} \underline{\epsilon}^{mn}{}_i = -  \epsilon^{mn}{}_{ik} n^k
\label{minus}.
\end{equation}
The minus appears because for timelike boundary the correct unit
normal for Stokes theorem is the inward pointing normal. Our $n^a$
is outward pointing. Substituting (\ref{minus}) into (\ref{iden}) we
get the required identity,
\begin{equation}
\underline{\epsilon}^{mn}{}_{(i} \underline{D}_{|m}
\pi_{n|j)} = - \frac{1}{2} \epsilon^{mn}{}_{ir}  C_{mnjs}n^s
n^r = - \beta_{ij} + {\cal
O}\left(\frac{1}{\rho}\right). \label{identity}
\end{equation}
 Upon expanding the left hand side of  (\ref{identity})
  using (\ref{pieqn}, \ref{c1}, \ref{c2}) we find
\begin{eqnarray}
\underline{\epsilon}^{mn}{}_i \underline{D}_m \pi_{nj} &=& \rho^2
\left( \sum_{p>0} \epsilon^{[p]mn}{}_i \rho^{-p}  \right) \left(
\sum_{q>0}D_m^{[q]} \rho^{-q}\right)
\left( \sum_{r>0}\pi^r_{nj} \rho^{-r}\right)   \nonumber \\
&=&  \Bigg{(}  \epsilon^{mn}{}_i D_m \pi^2_{nj} + \epsilon^{mn}{}_i
D_m^{[1]} \pi^1_{nj} + \epsilon^{mn}{}_i D_m^{[2]} \pi^0_{nj} +
\epsilon^{[1]mn}{}_i D_m \pi^1_{nj} + \epsilon^{[1]mn}{}_i D_m^{[1]}
\pi^0_{nj} \nonumber \\ & &+ \epsilon^{[2]mn}{}_i D_m \pi^0_{nj}
\Bigg{)} + {\cal O} \left( \frac{1}{\rho} \right)
 \nonumber \\
&=&   \epsilon_{mn(i} D^m\left(h^2_{j)}{}^n - 2 \s^2
\delta_{j)}^n\right)  + \mbox{ terms antisymmetric in $(i,j)$} + {\cal O}(\rho^{-1}). \label{beta2}
\end{eqnarray}
Combining (\ref{identity}) and (\ref{beta2}) we get the desired relation \be
\beta_{(ij)} = \epsilon_{mn(i} D^m\left(h^2_{j)}{}^n - 2 \s^2
\delta_{j)}^n\right)  + {\cal O}(\rho^{-1}).
\ee


\setcounter{equation}{0}
\section{A Collection of Useful Identities}

\label{collection}

Here we collect some identities which are often
used in the main text and in appendix \ref{pi} and
\ref{sectionidentity}. The Riemann tensor of the metric $h^0_{ab}$
on the unit hyperboloid ${\cal H}$ is given by
\begin{equation} R_{ijkl}^0 = (h_{ik}^0 h^0_{jl} - h_{jk}^0 h_{il}^0).  \end{equation}
Let $t, t_i $ and $t_{ij} = t_{(ij)}$ be some arbitrary fields on
the hyperboloid ${\cal H}$ then \cite{BS},
\begin{eqnarray}
\left[D_i, D^2\right] t       &=& -2 D_i t, \\
\left[D_i, D_j\right] t_k       &=& 2 h^0_{k[i} t_{j]}, \\
\left[D_i, D^2 \right]t_j  &= &  2 h^0_{ij} D_k t^k - 4 D_{(i}t_{j)},\\
\left[D_i, D_j\right] t_{k l} &=& 2 h^0_{i (k} t_{l)j} - 2 h^0_{j (k} t_{l)i}, \\
\left[D_i, D^2\right] t_{jk} &=& 4 h^0_{i(j} D^l t_{k)l} - 6 D_{(i}
t_{jk)}.
\end{eqnarray}
The following identities on $\s$ follows from commuting derivatives
and using the equation of motion for $\sigma$, i.e.,  (\ref{sigma})
\begin{eqnarray}
\s_{ik}{}^k  &=& - \s_i, \\
\s_{ijk}{}^k  &= & 6 \s h^0_{ij} + 3 \s_{ij}.
\end{eqnarray}
Using the expansion for $h_{ij}$ one finds the first and the second order Ricci tensors to be
\begin{eqnarray}
\label{ricci1}R^1_{ij} &=&
\s_{ij} - 3 \s h^0_{ij} \\ R^2_{ij} &=& - \frac{1}{2} D^2 h^2_{ij} -
\frac{1}{2} D_i D_j h^2 + D^k D_{(i}h^2_{j)k} + 3 \s_i \s_j + \s_k
\s^k h^0_{ij} + 2 \s \s_{ij} - 6 \s^2 h^0_{ij} \label{ricci2}
\end{eqnarray} and the Ricci scalar ($ \tilde R := \tilde h^{ij} R_{ij}$) is
\begin{eqnarray}
\label{ricciscalar}\tilde R &=& 6 + \frac{1}{\rho^2} \left( -2 h^2 - D^2
h^2 + D^i D^j h^2_{ij} + 6 \s_k \s^k - 24 \s^2 \right)  + {\cal
O}\left( \frac{1}{\rho^3}\right).
\end{eqnarray}
Finally we give first and second order correction to the
covariant derivative in terms of the connection
\begin{eqnarray}
 C^{[1]i}_{jk} &=&   - h_j^{0i} \sigma_k -
h^{0i}_k \sigma_j + \sigma^i h^0_{jk},  \label{c1}\\
C^{[2]i}_{jk} &=& - 2 \s \left( h^{0i}_j \s_k + h^{0i}_k \s_j -
h^0_{jk} \s^i \right) + \frac{1}{2} \left( D_j h^{2i}_k + D_k
h^{2j}_i - D^i h^2_{jk}\right) \label{c2}.
\end{eqnarray}



\begin{thebibliography}{99}


\bibitem{ADM1} R.~Arnowitt, S.~Deser and C.~W.~Misner, Nuovo Cimento,
\textbf{19} (1961) 668; R.~Arnowitt, S.~Deser and C.~W.~Misner, J.
Math. Phys. \textbf{1} (1960) 434; R.~Arnowitt, S.~Deser and
C.~W.~Misner, ``The Dynamics Of General Relativity,'' arXiv:gr-qc/0405109. 


\bibitem{RT} T.~Regge and C.~Teitelboim, ``Role Of Surface Integrals In The
Hamiltonian Formulation Of General Relativity,'' Annals Phys.\
\textbf{88},
286 (1974). 



\bibitem{AshtekarHansen} A. Ashtekar and R. O. Hansen, ``A Unified Treatment
of Null and Spatial Infinity in General Relativity. I. Universal
Structure, Asymptotic Symmetries, and Conserved Quantities at
Spatial Infinity,'' \textit{J. Math. Phys.}, \textbf{19} 1542
(1978).

\bibitem{AshtekarInf} A. Ashtekar, ``Asymptotic structure of the
gravitational field at spatial infinity'' in  \textit{General
relativity and gravitation : one hundred years after the birth of
Albert Einstein}, edited by A. Held (New York, Plenum Press, 1980).



\bibitem{AD}
  L.~F.~Abbott and S.~Deser,
  ``Stability Of Gravity With A Cosmological Constant,''
  Nucl.\ Phys.\ B {\bf 195}, 76 (1982).

\bibitem{RS}
 R.D.~Sorkin,
  ``Conserved Quantities as Action Variations'',
    in Isenberg, J.W., (ed.),
    {\it Mathematics and General Relativity}, pp. 23-37
    (Volume 71 in the AMS's Contemporary Mathematics series)
    (Proceedings of a conference, held June 1986 in Santa Cruz, California)
    (Providence, American Mathematical Society, 1988)


\bibitem{AR} A.~Ashtekar and J.~D.~Romano, ``Spatial infinity as
a boundary of space-time,'' Class.\ Quant.\ Grav.\ \textbf{9}, 1069
(1992).

 \bibitem{HH}   S.~W.~Hawking and G.~T.~Horowitz,
  ``The Gravitational Hamiltonian, action, entropy and surface terms,''
  Class.\ Quant.\ Grav.\  {\bf 13}, 1487 (1996)
  [arXiv:gr-qc/9501014].


\bibitem{SD2}
S.~Deser and B.~Tekin,
  ``Energy in generic higher curvature gravity theories,''
  Phys.\ Rev.\ D {\bf 67}, 084009 (2003)
  [arXiv:hep-th/0212292].


\bibitem{SD3}
  S.~Deser and B.~Tekin,
  ``Gravitational energy in quadratic curvature gravities,''
  Phys.\ Rev.\ Lett.\  {\bf 89}, 101101 (2002)
  [arXiv:hep-th/0205318].


\bibitem{Mann2}
R.~B.~Mann, ``Expanding the Area of Gravitational Entropy,''
Found.\ Phys.\  {\bf 33}, 65 (2003) [arXiv:gr-qc/0211047].




\bibitem{eeA} A.~Ashtekar, S.~Fairhurst and B.~Krishnan,
  ``Isolated horizons: Hamiltonian evolution and the first law,''
  Phys.\ Rev.\ D {\bf 62}, 104025 (2000)
  [arXiv:gr-qc/0005083].

\bibitem{ABL}
 A.~Ashtekar, C.~Beetle and J.~Lewandowski,
  ``Mechanics of Rotating Isolated Horizons,''
  Phys.\ Rev.\ D {\bf 64}, 044016 (2001)
  [arXiv:gr-qc/0103026].



 \bibitem{MM}
  R.~B.~Mann and D.~Marolf,
``Holographic renormalization of asymptotically flat spacetimes,''
  Class.\ Quant.\ Grav.\  {\bf 23}, 2927 (2006)
  [arXiv:hep-th/0511096].






\bibitem{Mann} R.~B.~Mann, ``Misner string entropy,'' Phys.\ Rev.\ D \textbf{%
60}, 104047 (1999) [arXiv:hep-th/9903229]. 

\bibitem{KLS} P.~Kraus, F.~Larsen and R.~Siebelink, ``The gravitational
action in asymptotically AdS and flat spacetimes,'' Nucl.\ Phys.\ B \textbf{%
563}, 259 (1999) [arXiv:hep-th/9906127]. 





\bibitem{HIM2} S.~Hollands, A.~Ishibashi and D.~Marolf, ``Counter-term
charges generate bulk symmetries,'' Phys.Rev. {\bf D72} 104025
(2005) [arXiv:hep-th/0503105].

\bibitem{Peierls} R. E. Peierls, ``The commutation laws of relativistic
field theory,'' Proc. Roy. Soc. (London) {\bf 214} (1952),
143-157.

\bibitem{BY} J.~D.~Brown and J.~W.~.~York, ``Quasilocal energy and conserved
charges derived from the gravitational action,'' Phys.\ Rev.\ D \textbf{47},
1407 (1993). 


\bibitem{skenderis} M.~Henningson and K.~Skenderis, ``The holographic Weyl
anomaly,'' JHEP \textbf{9807}, 023 (1998) [arXiv:hep-th/9806087].

\bibitem{Kraus} V. ~Balasubramanian and P.~Kraus, ``A stress tensor for
anti-de Sitter gravity,'' Commun.\ Math.\ Phys.\ \textbf{208}, 413 (1999)
[arXiv:hep-th/9902121].


\bibitem{HIM}
S.~Hollands, A.~Ishibashi and D.~Marolf, ``Comparison between
various notions of conserved charges in  asymptotically
  AdS-spacetimes,''
  Class.\ Quant.\ Grav.\  {\bf 22}, 2881 (2005)
  [arXiv:hep-th/0503045].


\bibitem{KSnew}
 I.~Papadimitriou and K.~Skenderis,
  ``Thermodynamics of asymptotically locally AdS spacetimes,''
JHEP {\bf 0508}, 004 (2005)   [arXiv:hep-th/0505190.]

\bibitem{AV}
 A.~Ashtekar and M.~Varadarajan,
  ``A Striking property of the gravitational Hamiltonian,''
  Phys.\ Rev.\ D {\bf 50}, 4944 (1994)
  [arXiv:gr-qc/9406040].

\bibitem{DonLeo}
  D.~Marolf and L.~Patino,
 ``The non-zero energy of 2+1 Minkowski space,''
 Phys.Rev. {\bf D74} 024009 (2006)
  [arXiv:hep-th/0604127].



\bibitem{BS} R. Beig and B. Schmidt, ``Einstein's equations near spatial
infinity,'' 1982 Commun. Math. Phys. 87 65.




\bibitem{B} R. Beig, ``Integration of Einstein's Equations Near Spatial
Infinity,'' 1984 Proc. R. Soc. A 391 295.



\bibitem{Wald}
R.~M.~Wald, ``General Relativity,'' The University of Chicago Press,
1984.




\bibitem{BCM}
  J.~D.~Brown, J.~Creighton and R.~B.~Mann,
  ``Temperature, energy and heat capacity of asymptotically anti-de Sitter
  black holes,''
  Phys.\ Rev.\ D {\bf 50}, 6394 (1994)
  [arXiv:gr-qc/9405007].





\bibitem{DAR}
D.~Astefanesei and E.~Radu, ``Quasilocal formalism and black ring
thermodynamics,'' Phys.Rev. {\bf D73} 044014 (2006)
[arXiv:hep-th/0509144.]



\bibitem{KKmonopole}
R.B. Mann and C. Stelea, ``On the gravitational energy of the
Kaluza Klein monopole", Phys.Lett. {\bf B634}  531 (2006)
[arXiv:hep-th/0511180]



\bibitem{ABR} A.~Ashtekar, L.~Bombelli and O.~Reula, ``The Covariant Phase
Space Of Asymptotically Flat Gravitational Fields,'' in
\textit{Analysis, Geometry and Mechanics: 200 Years After Lagrange},
edited by M. Francaviglia and D. Holm (North-Holland, Amsterdam,
1991).



\bibitem{skenAF} S.~de Haro, K.~Skenderis and S.~N.~Solodukhin,
``Gravity in warped compactifications and the holographic stress
tensor,'' Class.\ Quant.\ Grav.\  {\bf 18}, 3171 (2001)
[arXiv:hep-th/0011230].





\bibitem{AM1} A. Ashtekar and A. Magnon-Ashtekar, ``Energy-Momentum in General Relativity,''  Phys. Rev. Lett. \textbf{43}, 181-184
(1979), Erratum: Phys. Rev. Lett. \textbf{43}, 649 (1979).

\bibitem{AM2}A.~Ashtekar and A.~Magnon, ``From $i^0$ to
the 3 + 1 description of spatial infinity,''\textit{J. Math. Phys.},
\textbf{25} 2682 (1984).


\bibitem{AshLog} A. Ashtekar, ``Logarithmic Ambiguities in the
Description of Spatial Infinity,'' Found.\ Phys.\ {\bf 15}, 419
(1985).







\end{thebibliography}
\end{document}